\pdfoutput=1
\documentclass[a4paper,american,citeautoscript,floatfix,pdftex,superscriptaddress,%
aps,pra,%
longbibliography,%
]{revtex4-2}%

\usepackage{graphicx}
\usepackage{amssymb,amsfonts,dsfont}
\usepackage{tikz}
\usetikzlibrary{quantikz2}
\usepackage[T1]{fontenc}
\usepackage[utf8]{inputenc}
\usepackage{microtype}
\usepackage{xspace}
\usepackage{hyperref, hypernat}

\usepackage[paper=portrait,pagesize]{typearea}

\usepackage{array}
\usepackage{multirow}
\usepackage{tabularray}
\usepackage{amsmath} 
\usepackage{tabularx,ragged2e,booktabs}
\usepackage{rotating}
\usepackage{mathtools}
\usepackage{booktabs}

\hypersetup{citebordercolor=yellow,linkbordercolor=red,urlbordercolor=blue} %

\newcommand{\beit}{\affiliation{BEIT Inc., Wadowicka 8A 30-415 Kraków, Poland}}%
\newcommand{\beitemail}{\email[]{emil@beit.tech}}%
\newcommand{\beitweb}{\homepage{https://www.beit.tech}}%

\begin{document}

\title{Fault-tolerant quantum simulation of the Pauli-Breit Hamiltonian for ab initio hybrid quantum-classical molecular design with applications to photodynamic therapy}%
\author{Emil Zak}\beit\beitemail\beitweb

\begin{abstract}
	Relativistic spin effects are driving subtle molecular processes ranging from intersystem crossing in photodynamic therapy to spin-mediated catalysis and high-resolution spectroscopy. These effects can be described by the Pauli-Breit Hamiltonian, which extends the nonrelativistic electronic Hamiltonian by including explicit one- and two-electron spin-orbit and spin-spin interactions. However, first-principles simulations of the full Pauli-Breit Hamiltonian quickly becomes intractable on classical computers due to the rapid growth of Hilbert space dimension and the complexity of two-body spin-dependent terms.
	
	In this work, we propose a fault-tolerant quantum algorithm for computing molecular energy levels and properties governed by the Pauli-Breit Hamiltonian. Our approach is based on block-encoding of the relativistic Hamiltonian in a second-quantized, doubly factorized representation. By reformulating the Pauli-Breit Hamiltonian in a symmetry-adapted Majorana basis, we construct efficient linear-combination-of-unitaries circuits that encode both one- and two-electron spin-orbit coupling without resorting to effective or mean-field approximations. We introduce spin-controlled Pauli-SWAP networks that decouple spin and orbital control logic, enabling a unified treatment of relativistic spin mixing with only a modest overhead relative to spin-free electronic structure simulations.
	
	We analyze the resulting quantum resources in terms of logical qubit counts, T-gate complexity, and show that the inclusion of spin degrees of freedom does not fundamentally worsen the asymptotic scaling. The prefactor in our approach is 2x lower than if the linear-combination of unitaries technique was applied directly. 
		
	To demonstrate the prospects for using the algorithm, we propose a first-principles hybrid quantum-classical workflow for the rational design of photodynamic therapy photosensitizers, artificial photosynthesis catalysts, and other molecular systems where accurate treatment of relativistic spin effects is essential.
\end{abstract}

\maketitle

\section{Introduction}
\label{sec:introduction}

The Pauli-Breit (PB) Hamiltonian is a perturbative approximation to the relativistic Dirac equation that provides a systematic extension of the non-relativistic electronic Hamiltonian by including explicit spin-dependent interactions~\cite{Breit1932,Foldy1950}. In addition to scalar relativistic corrections, the PB Hamiltonian includes one- and two-body spin-orbit and spin-spin coupling terms, describing relativistic spin effects within a second-quantized many-electron formulation~\cite{Gould1993,He1996}.

Relativistic effects play an essential role in quantum chemistry whenever heavy elements are present, or when spin mixing between electronic states of different multiplicity is relevant, such as in intersystem crossing processes. Even for molecules composed of relatively light atoms, accurate treatment of spin-orbit coupling can be required when high spectroscopic precision is sought or when electronic states are nearly degenerate. However, first-principles calculations that include the full PB Hamiltonian together with high-quality basis sets remain beyond the capabilities of present-day classical computers, even for small molecular systems~\cite{Shiozaki2015,Sun2022,Tantardini2024}. This computational barrier sets in mainly because coupling different non-relativistic electronic states causes the computation volume to increase several-fold, compared to electronic structure calculations with the Schroedinger equation. As a consequence, most practical approaches rely on truncated or effective treatments of relativistic spin interactions~\cite{He1996,Penfold2018}.

At various levels of approximation, established quantum chemistry packages such as CFOUR~\cite{cfour}, DIRAC~\cite{DIRAC24}, MOLCAS and OpenMolcas~\cite{LiManni2023}, and MOLPRO~\cite{Werner2020} support calculations that include selected components of the PB Hamiltonian. While these approaches have proven successful in many applications, they typically involve approximations whose validity depends on the system and the desired accuracy~\cite{Valentine2022,Penfold2018,Hagai2024,Epifanovsky2015,BERNING2000,Aquilante2015}. Fault-tolerant quantum computing (FTQC) architectures are anticipated to provide a route toward eliminating these approximations and enabling fully first-principles simulations of relativistic electronic structure.

In this work, we propose a quantum algorithm for simulating the Pauli-Breit Hamiltonian on fault-tolerant quantum computers. Our approach is based embedding the Hamiltonian into a unitary quantum circuit, followed by the use of quantum phase estimation (QPE)~\cite{babbush2018} for calculating energy levels. We achieve this embedding using the block-encoding framework~\cite{Low2019}, in which the Hamiltonian appears as a submatrix of a larger unitary operator. Within this paradigm, the overall quantum resource requirements are determined by the target precision and by the cost of implementing the block-encoding circuit. Throughout this work, we quantify quantum resources primarily in terms of logical qubit counts and T-gate complexity.

Our main technical contribution is the construction of block-encoding circuits for the Pauli-Breit Hamiltonian expressed in a second-quantized, doubly factorized representation~\cite{zak2024}. Building on the framework introduced by von Burg \textit{et al.}~\cite{Burg2021} for non-relativistic electronic Hamiltonians, we extend this approach to include spin-mixing operators arising from both one- and two-body spin-orbit coupling terms. This extension gives a unified description of relativistic spin effects within a fault-tolerant quantum algorithmic setting.

High-accuracy quantum chemistry calculations that include relativistic spin effects are required across a broad range of applications in physics, chemistry, and materials science. Spin-orbit and spin-spin interactions play a central role in relaxation mechanisms in photomaterials~\cite{Rose2014,MuozCastro2023,Hagai2024}, controlled chemical reactions~\cite{Recio2022}, reactive oxygen species generation rates in photosensitizers for photodynamic therapy~\cite{Knoll2015,Yang2024}, enzymatic reactions~\cite{Minaev2017}, oxygen binding in heme systems~\cite{Vickers2006}, and core-level spectroscopies~\cite{Jayadev2025}. 
In high-resolution spectroscopy, such as studies of the CH$_2$ radical, accurate treatment of singlet-triplet interactions is essential for describing collision-induced intersystem crossing and related processes~\cite{Vahtras1992,Osmann1997,Bouallagui2023}. Beside spin-orbit interactions governing the generation of reactive oxygen species in photodynamic therapy~\cite{Knoll2015,Yang2024}, they steer artificial photosynthesis, for example in ruthenium-based photocatalysts for CO$_2$ conversion~\cite{Konrath2020,alamos,Mai2017,Wang2021}. Moreover, spin-spin interactions are indispensable for quantitative simulations of EPR and NMR spectra.
Being able to model these interactions substantially enlarges the class of systems and phenomena that can be addressed compared with simulations based solely on the non-relativistic electronic Hamiltonian.

From a computational perspective, the main challenge in simulating the Pauli-Breit Hamiltonian is due the two-body spin-orbit coupling terms~\cite{Fedorov2003,Penfold2018,BERNING2000}. Although these terms are sometimes neglected or absorbed into effective one-body operators due to partial error cancellation~\cite{He1996}, such approximations can fail when high accuracy is required. For instance, in formaldehyde the explicit inclusion of two-electron spin-orbit coupling changes excitation energies by approximately 10~cm$^{-1}$ relative to effective one-body Hamiltonian~\cite{Kotaru2022}, while in PsoralenOS the two-electron contribution can reach up to 100\% of the one-electron term~\cite{Kotaru2022}. 
Thus while the one-electron contribution to spin-orbit coupling often dominates, the two-electron terms can be significant and, in some cases, comparable in magnitude. Further examples include the two-electron spin-orbit contribution amounting to approximately 10\% of the total effect in SeH, about 19\% in SH and 35\% in OH, showing its relevance even in light-element systems~\cite{Valentine2022}. 
These examples highlight the need for a unified and systematically improvable model of both one- and two-body spin-orbit interactions~\cite{MuozCastro2023}.

The Hilbert space size for a complete active space calculation for spin $S$ system with $N$ orbitals and $\eta$ electrons is given by:
\begin{eqnarray}
	N_S = \frac{2S+1}{N+1}\binom{N+1}{\frac{1}{2}\eta-S}\binom{N+1}{N-\frac{1}{2}\eta-S}
	\label{eq:Ns}
\end{eqnarray}
As a consequence of eq.~\ref{eq:Ns}, the inclusion of spin degrees of freedom does not exacerbate the curse of dimensionality in the same manner as increasing the number of electrons or spatial orbitals. Under field-free conditions, the spin degree of freedom introduces a multiplicative factor of $(2S+1)$ in the size of the Hilbert space for a fixed total spin $S$. Consequently, the additional cost associated with spin mixing enters as a controlled overhead. We show that, within the block-encoding framework developed here, this overhead can be managed such that the asymptotic scaling remains close to that of the spin-free electronic Hamiltonian.

Electronic energy levels can be computed using the quantum phase estimation (QPE) algorithm, which is widely regarded as one of the most efficient approaches for eigenvalue estimation on fault-tolerant quantum computers~\cite{babbush2018}. QPE extracts eigenvalues by accumulating the phase generated by the time-evolution operator $e^{iHt}$, assuming access to a suitable approximate eigenstate and a target precision $\varepsilon$.
Efficient implementations of the time-evolution operator can be achieved using  Trotterization, quantum signal processing (QSP)~\cite{low2017}, and qubitization~\cite{Low2019}. In particular, QSP- and qubitization-based realizations of QPE show favorable asymptotic complexity, requiring $\mathcal{O}(\zeta t + \log(1/\varepsilon))$ applications of a unitary Hamiltonian oracle $\mathcal{B}[H]$ that encodes the scaled Hamiltonian $H/\zeta$.
Consequently, the total number of oracle calls required by QPE scales linearly with the block-encoding normalization factor, $\mathcal{O}(\zeta)$. 

On the quantum algorithmic side, resource estimates for block-encoded nonrelativistic electronic Hamiltonians have been extensively studied~\cite{babbush2018,Burg2021,lee2021,rocca2024,loaiza2024,Deka2025}. One route to reducing quantum resource requirements lies in choosing representations that lower the normalization factor $\zeta$ of the encoded Hamiltonian, either through basis selection or factorization schemes. In this work, we adopt the doubly factorized representation of Ref.~\cite{Burg2021}, while noting that alternative approaches such as tensor hypercontraction~\cite{lee2021} may offer further advantages. During the preparation of this manuscript, a notable work was published in Ref.~\cite{Xanadu25} on fault-tolerant quantum simulations of spin-orbit interactions, with applications to photodynamic therapy. That study presents detailed resource estimates and algorithmic advances that are to some extent complementary to, but not covered in, the present work. By contrast, our study focuses on handling spin mixing in the Pauli-Breit Hamiltonian and on outlining prospects for applications within hybrid quantum-classical schemes.

\section{The Pauli-Breit Hamiltonian}
The Pauli-Breit Hamiltonian for an $\eta$-electron, $M$-nuclei molecule is given as~\cite{Gould1993,Kent1994}:
\begin{equation}
	\hat{H}_{PB}=\hat{H}^{(el.)}+\hat{H}^{(D1)}+\hat{H}^{(D2)}+\hat{H}^{(SO,1)}+\hat{H}^{(SO,2)}+\hat{H}^{(SS)}
	\label{eq:ham-PB}
\end{equation}
$\hat{H}^{(el.)}$ is the nonrelativistic electronic Hamiltonian within the Born-Oppenheimer approximation written as
\begin{equation}
	\hat{H}^{(el.)} = \frac{1}{2}\sum_{j=1}^{\eta}\hat{\mathbf{p}}_j^2-\sum_{j=1}^{\eta}\sum_{\alpha=1}^{M}\frac{Z_{\alpha}}{||\mathbf{r_j}-\mathbf{R_{\alpha}}||}+\sum_{j<k}^{\eta}\frac{1}{||\mathbf{r_j}-\mathbf{r_k}||}
	\label{eq:ham-el}
\end{equation}
where $\mathbf{\hat{p}}_j$ is the momentum operator for the $j$-th electron, $Z_{\alpha}$ is the nuclear charge of atom $\alpha$, $\mathbf{\hat{r}}_j \in \mathbb{R}^3$ is the position operator of the $j$-th electron and  $\mathbf{R_{\alpha}} \in \mathbb{R}^3$ is the position for the $\alpha$-th nucleus. Atomic units are used throughout this paper and terms beyond the Born-Oppenheimer approximation have been neglected. The terms written in eq.~\ref{eq:ham-el} are respectively: the total electronic kinetic energy $\hat{T}^{(e)}$, the nuclear-electron attraction  $\hat{V}^{(ne)}$  and the electron-electron interaction $\hat{V}^{(ee)}$. 
In eq.~\ref{eq:ham-PB}, $\hat{H}^{(D1)}$ is the one-electron Darwin term and the kinetic energy associated with the relativistic electronic mass-increase, respectively:
\begin{equation}
	\hat{H}^{(D1)} = \frac{\pi}{2}\sum_{j=1}^{\eta}\sum_{\alpha=1}^{M}Z_{\alpha}\delta\left(\mathbf{\hat{r}}_j-\mathbf{R_{\alpha}}\right)-\frac{1}{8}\sum_{j=1}^{\eta}\hat{\mathbf{p}}_j^4.
	\label{eq:ham-D1}
\end{equation}
$\hat{H}^{(D2)}$ includes the electron orbit-orbit interaction and the relativistic two-electron Darwin term, respectively:
\begin{equation}
	\hat{H}^{(D2)} =  -\frac{1}{2}\sum_{j<k}^{\eta}\left(\hat{\mathbf{p}}_j\cdot \left[\frac{(\mathbf{\hat{r}}_j-\mathbf{\hat{r}}_k)(\mathbf{\hat{r}}_j-\mathbf{\hat{r}}_k)}{||\mathbf{\hat{r}}_j-\mathbf{\hat{r}}_k||^3} +\frac{1}{||\mathbf{\hat{r}}_j-\mathbf{\hat{r}}_k||}\right]\cdot \hat{\mathbf{p}}_k - \pi\delta\left(\mathbf{\hat{r}}_j-\mathbf{\hat{r}}_k\right)  \right)
	\label{eq:ham-D2}
\end{equation}
Hamiltonians $\hat{H}^{(el.)}$,$\hat{H}^{(D1)}$, and $\hat{H}^{(D2)}$ form the spin-independent part of the Pauli-Breit Hamiltonian. The spin-dependent part carries contributions from the spin-orbit couplings $\hat{H}^{(SO,1)}, \hat{H}^{(SO,2)}$ and the spin-spin interaction $\hat{H}^{(SS)}$. The \textit{one-body} spin-orbit coupling (SOC) is given as:
\begin{equation}
	\hat{H}^{(SO,1)} =  \frac{1}{2}\sum_{j=1}^{\eta}\sum_{\alpha=1}^{M}\frac{Z_{\alpha}}{||\hat{\mathbf{r}}_j-\mathbf{R_{\alpha}}||^3}\hat{\mathbf{s}}_j\cdot\left[\hat{\mathbf{l}}_j-(\mathbf{R_{\alpha}}\times \hat{\mathbf{p}}_j)\right]
	\label{eq:ham-SO1}
\end{equation}
where $\hat{\mathbf{s}}_j$ is the spin operator for the $j$-th electron and $\hat{\mathbf{l}}_j=\hat{\mathbf{r}}_j \times \hat{\mathbf{p}}_j$ is the angular momentum operator for the $j$-th electron. The one electron spin-orbit interaction term $\hat{H}^{(SO,1)}$ given in eq.~\ref{eq:ham-SO1} describes the interaction between electron's spin with the magnetic moment created by its orbital motion in the  electrostatic field of nuclei. The \textit{two-body} spin orbit coupling $\hat{H}^{(SO,2)}$ is given by:
\begin{equation}
	\hat{H}^{(SO,2)} = \sum_{j<k}^{\eta}\frac{(\hat{\mathbf{s}}_j+\hat{\mathbf{s}}_k)}{||\hat{\mathbf{r}}_j-\hat{\mathbf{r}}_k||^3}\cdot\left[(\hat{\mathbf{r}}_j-\hat{\mathbf{r}}_k)\times (\hat{\mathbf{p}}_j-\hat{\mathbf{p}}_k)\right]
	\label{eq:ham-SO2}
\end{equation}
The two-electron spin-orbit interaction term $\hat{H}^{(SO,2)}$ given in eq.~\ref{eq:ham-SO2} 
is due to the interaction of electron's spin with the orbital magnetic moment of another electron. $\hat{H}^{(SO,2)}$ is often expressed in a more compact form written as:
\begin{equation}
	\hat{H}^{(SO,2)} = \sum_{j<k}^{\eta}\frac{1}{||\hat{\mathbf{r}}_j-\hat{\mathbf{r}}_k||^3}\hat{\mathbf{l}}_{jk}\cdot\left[\hat{\mathbf{s}}_j+2\hat{\mathbf{s}}_k\right]
	\label{eq:ham-SO2-alt}
\end{equation}
where $\hat{\mathbf{l}}_{jk}=(\hat{\mathbf{r}}_j-\hat{\mathbf{r}}_k)\times \hat{\mathbf{p}}_j$. For molecules composed solely of light elements, the two-electron SOC term $\hat{H}^{(SO,2)}$ can be non-negligible \cite{He1996}. In practice, this contribution is often treated within a mean-field approximation, reducing the two-electron SOC operator to an effective one-electron form \cite{He1996,Saue2011}. For systems containing heavy elements, the one-electron SOC term typically dominates, and the two-electron contribution is therefore frequently neglected \cite{Saue2011,Penfold2018,Fedorov2003} or expressed through an effective formulations. Within a one-center approximation applied to the full spin-orbit Hamiltonian $\hat{H}^{(SO,1)} + \hat{H}^{(SO,2)}$, one obtains an effective operator of the form $\hat{H}^{(SO,eff)}=\sum_{j=1}^{\eta}\sum_{\alpha=1}^M\frac{Z^{eff}_{\alpha}}{r^3_{j\alpha}}\hat{\mathbf{l}}_j\hat{\mathbf{s}}_j=\sum_{j=1}^{\eta}\xi(\hat{\mathbf{r}}_j)\hat{\mathbf{l}}_j\hat{\mathbf{s}}_j$, where $\xi(\mathbf{r}_j) = \sum_{\alpha=1}^{M} Z^{\mathrm{eff}}_{\alpha}/r^{3}_{j\alpha}$ is the effective spin-orbit coupling constant. While such approximations are widely used~\cite{He1996}, they should be avoided in high-accuracy calculations, where explicit treatment of both one- and two-electron SOC terms is required \cite{Penfold2018,Fedorov2003}. This paradigm assumes first-principle calculations maximally avoiding approximations, to maintain a better error control.

\begin{figure}[!h] 
	\centering
	\includegraphics[width=0.3\linewidth]{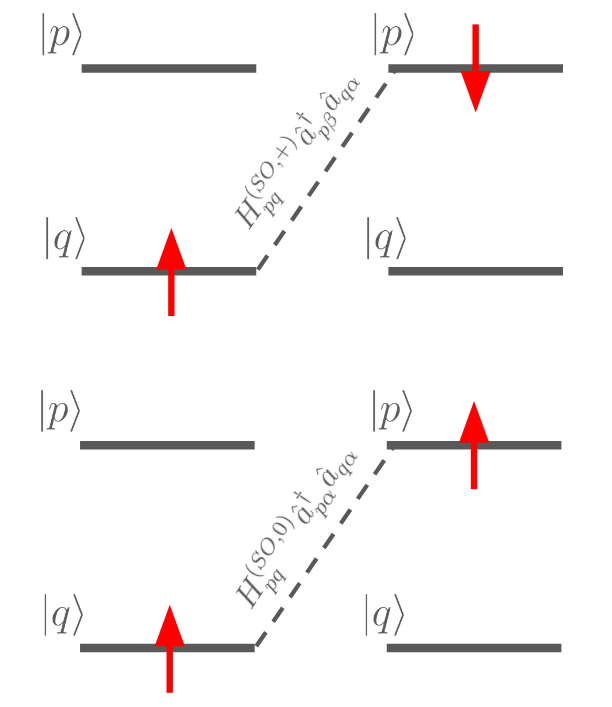}
	\caption{Spin-orbit induced electronic excitations between spin-orbitals and the respective coupling matrix elements. }
	\label{fig:excitation}
\end{figure}

The $\hat{H}^{(SS)}$ term given in eq.~\ref{eq:ham-PB} includes \textit{spin-spin} electron interactions and the \textit{contact term}, respectively:
\begin{equation}
	\hat{H}^{(SS)} = \sum_{j<k}^{\eta}\frac{1}{||\hat{\mathbf{r}}_j-\hat{\mathbf{r}}_k||^3}\hat{\mathbf{s}}_j\cdot \left[3\frac{(\hat{\mathbf{r}}_j-\hat{\mathbf{r}}_k)(\hat{\mathbf{r}}_j-\hat{\mathbf{r}}_k)}{||\hat{\mathbf{r}}_j-\hat{\mathbf{r}}_k||^2} -1 \right]\cdot \hat{\mathbf{s}}_k - \frac{8}{3}\pi\delta\left(\hat{\mathbf{r}}_j-\hat{\mathbf{r}}_k\right)\hat{\mathbf{s}}_j\hat{\mathbf{s}}_k .
	\label{eq:ham-SS}
\end{equation}

\subsection{Symmetry-adapted Pauli-Breit Hamiltonian in second-quantization}
The PB Hamiltonian can be represented in a basis of $2N$ orthonormal molecular spin-orbitals:
\begin{equation}
	B_N = \lbrace {\ket{p\sigma}}\;|\; p=1,...,N; \sigma = -\frac{1}{2},+\frac{1}{2} \rbrace 
	\label{eq:so-basis}
\end{equation}
where index $p$ labels the orbitals and $\sigma$  denotes the spin. The spin-orbital basis functions can be expressed in the position representation as follows:
\begin{equation}
	\braket{\mathbf{r},\xi}{p\sigma} = \psi_{p\sigma}(\mathbf{r},\xi)  = \phi_{p}(\mathbf{r})\cdot\chi_{\sigma}(\xi) 
	\label{eq:so-wf}
\end{equation}
where $\ket{r}$ and $\ket{\xi}$ are eigenvector of the electron position operator and the spin standard basis, respectively. 
In second-quantization, the PB Hamiltonian is represented with fermionic creation and annihilation operators $\hat{a}^{\dagger}_{p\sigma}$ and $\hat{a}_{p\sigma}$, respectively. These spin-orbital ladder operators satisfy the fermionic anticommutation relation: $\left[\hat{a}^{\dagger}_{p\sigma}, \hat{a}_{q\rho}\right]_{+}=\delta_{pq}\delta_{\sigma\rho}$, where $\delta_{pq}$ is the Kronecker's delta.
The thus formed finite-dimensional Hilbert space $\mathcal{H}(2N)$ spans irreducible representations of the $\mathcal{U}(2N)$ Lie group with generators~\cite{Gould1993} given as:
\begin{equation}	
	\hat{E}_{pq,\sigma\rho}=\hat{a}^{\dagger}_{p\sigma}\hat{a}_{q\rho}
	\label{eq:U(2N)-gen}
\end{equation}
With such a representation, the one-electron operators are written as:
\begin{equation}
	\hat{H}^{(1)}= \sum_{p,q=1}^N\sum_{\sigma,\rho = \pm \frac{1}{2}}\bra{p,\sigma}	\hat{H}^{(1)} \ket{q,\rho}\hat{E}_{pq,\sigma\rho}
	\label{eq:one-el}
\end{equation}
and the two-electron operators:
\begin{equation}
	\hat{H}^{(2)}= \frac{1}{2}\sum_{p,q,r,s=1}^N\sum_{\sigma,\rho, \tau,\nu = \pm \frac{1}{2}}\bra{pq,\sigma\rho}	\hat{H}^{(2)} \ket{rs,\tau\nu}\left(\hat{E}_{pr,\sigma\tau}\hat{E}_{qs,\rho\nu} - \delta_{qr}\delta_{\rho\tau}\hat{E}_{ps,\sigma\nu}\right)
	\label{eq:two-el}
\end{equation}
where
\begin{equation}
	\bra{pq,\sigma\rho}\hat{H}^{(2)}\ket{rs,\tau\nu}
	=
	\iint_{\mathbb{R}^3\times\mathbb{R}^3}
	d\mathbf r_1\, d\mathbf r_2
	\sum_{\xi_1,\xi_2}
	\psi^{*}_{p\sigma}(\mathbf r_1,\xi_1)\,
	\psi^{*}_{q\rho}(\mathbf r_2,\xi_2)\,
	\hat{H}^{(2)}(\mathbf r_1,\xi_1;\mathbf r_2,\xi_2)\,
	\psi_{r\tau}(\mathbf r_1,\xi_1)\,
	\psi_{s\nu}(\mathbf r_2,\xi_2).
\end{equation}

The spin-independent operators $\hat{H}^{(el.)},\hat{H}^{(D1)},\hat{H}^{(D2)}$ are expressible in term of the $\mathcal{U}(N)$ generators:
\begin{equation}
	\hat{E}_{pq}=\sum_{\sigma}\hat{E}_{pq,\sigma\sigma}
	\label{eq:U(N)-gen}
\end{equation}
which commute with the $\mathcal{U}(2)$ spin group generators:
\begin{equation}	
	\hat{E}_{\sigma\rho}=\sum_{p}\hat{E}_{pp,\sigma\rho}
	\label{eq:U(2)-gen}
\end{equation}
For spin-independent problems, it is therefore sufficient to work within a given irreducible representation of the group $\mathcal{U}(N)$, corresponding to the unitary group of spatial orbitals \cite{Gould1993}. In contrast, for spin-dependent problems, such as those involving spin-orbit coupling, the operators defined in Eqs.~\eqref{eq:U(N)-gen} and \eqref{eq:U(2)-gen} generate a $\mathcal{U}(N)\times\mathcal{U}(2)$ subgroup of $\mathcal{U}(2N)$. Consequently, it is sufficient to operate in a $\mathcal{U}(N)\times\mathcal{U}(2)$ symmetry-adapted spin-orbital basis, i.e. a product basis.

The one-body part of the Pauli-Breit Hamiltonian can be expressed as follows:
\begin{equation}
	\hat{H}^{(1)} = \sum_{\mu = X,Y,Z,0} \sum_{\sigma,\rho} \sum_{p,q}\hat{P}^{(\mu)}_{\sigma\rho}H^{(1,\mu)}_{pq}\hat{E}_{p\sigma q\rho}
	\label{eq:one-body-second-quant}
\end{equation}
where 
\begin{eqnarray}
	H^{(1,\mu)}_{pq} = \bra{p} \delta_{\mu 0}\left(\hat{T}^{(e)}+\hat{V}^{(ne)}+\hat{H}^{(D1)}\right)+\frac{1}{2}\left(1-\delta_{\mu 0}\right)\hat{H}^{(orb,1,\mu)}\ket{q}.
	\label{eq:ham-1}
\end{eqnarray}
with detailed derivation given in Appendix~\ref{sec:AppendixA}.
The Hamiltonian shown in eq.~\ref{eq:one-body-second-quant} can be rewritten as the scalar product of the following vectors:
\begin{eqnarray}
	\mathbf{H^{(1)}} = \begin{pmatrix}
		\hat{T}^{(e)}+\hat{V}^{(ne)}+\hat{H}^{(D1)} \\
		\hat{H}^{(orb,1,X)}\\
		\hat{H}^{(orb,1,Y)}\\
		\hat{H}^{(orb,1,Z)}
	\end{pmatrix}, \; \mathbf{\hat{P}}=\begin{pmatrix}
	\mathbf{1} \\
	\hat{X}\\
	\hat{Y}\\
	\hat{Z}
	\end{pmatrix}
\end{eqnarray}
giving:
\begin{equation}
	\hat{H}^{(1)} = \sum_{p,q}\sum_{\sigma,\rho}\mathbf{H}^{(1)}_{pq}\cdot\mathbf{\hat{P}}_{\sigma\rho}\hat{E}_{p\sigma q\rho}
	\label{eq:one-body-part}
\end{equation}
where $\mathbf{\hat{P}}_{\sigma\rho}$ are elements of Pauli matrices $\mathbf{1},\hat{X},\hat{Y},\hat{Z}$. We dropped orbital and spin summation ranges for clarity of presentation. Note that $\hat{E}_{pq} = \sum_{\sigma,\rho}\mathbf{1}_{\sigma\rho}\hat{E}_{p\sigma q\rho}$. The spin-other-orbit, electron-electron, Darwin and spin-spin interactions contain one-body terms, which we collate in the form of an \textit{add-on} Hamiltonian:
\begin{equation}
	\begin{split}
	\hat{H}^{(1,add-on)} = \sum_{\mu,\mu'=0,X,Y,Z}\sum_{p,q,r,s}\sum_{\sigma,\rho, \tau,\nu}\left[\delta_{\mu 0}\delta_{0\mu'}\hat{P}^{(0)}_{\sigma\rho}\hat{P}^{(0)}_{\tau\nu}\left(\hat{E}_{r\tau s \nu}\frac{1}{2}\bra{pq}\hat{V}^{(ee)}+\hat{H}^{(D2)}\ket{rs}-\right.\right.\\
	\left.\left.-\hat{E}_{r\tau s \nu}\frac{1}{4}\bra{pq}\hat{H}^{(cont)}\ket{rs}\right)+\right.\\
	\left.+(1-\delta_{0\mu'}-\delta_{\mu 0}+\delta_{\mu 0}\delta_{0\mu'})\hat{P}^{(\mu)}_{\sigma\rho}\hat{P}^{(\mu')}_{\tau\nu}\hat{E}_{r\tau s \nu}\frac{1}{2}\bra{pq}\hat{H}^{(SS,\mu\mu')}\ket{rs}+\right.\\
	\left. +\delta_{0\mu'}(1-\delta_{\mu 0})\hat{P}^{(\mu)}_{\sigma\rho}\hat{P}^{(\mu')}_{\tau\nu}\hat{E}_{r\tau s \nu}\frac{1}{2}\bra{pq}\hat{H}^{(orb,2,\mu)}\ket{rs}+\right.\\
	\left. +\delta_{0\mu}(1-\delta_{\mu' 0})\hat{P}^{(\mu)}_{\sigma\rho}\hat{P}^{(\mu')}_{\tau\nu}\hat{E}_{q\sigma s \rho}\frac{1}{2}\bra{pq}\hat{H}^{(orb,2,\mu')}\ket{rs}
	\right]	
	\label{eq:one-body-addon}
    \end{split}
\end{equation}
Combination of eq.~\ref{eq:ham-1} with eq.~\ref{eq:one-body-addon} gives a one-body operator, written in a general form as:
\begin{equation}
	\hat{H}^{(1)} = \sum_{p,q}\sum_{\sigma,\rho}\mathbf{\tilde{H}}^{(1)}_{pq}\cdot\mathbf{\hat{P}}_{\sigma\rho}\hat{E}_{p\sigma q\rho}
	\label{eq:one-body-full}
\end{equation}
where $\mathbf{\tilde{H}}^{(1)}$ represents a Hamiltonian constructed by combining the Hamiltonian given in eqs.~\ref{eq:one-body-second-quant},~\ref{eq:one-body-addon}.
The Hamiltonian shown in eq.~\ref{eq:one-body-full} can be further expressed with \textit{extended triplet excitation operators} defined as:
\begin{eqnarray}
	\hat{\tilde{T}}_{pq}^{(\mu)} = \sum_{\sigma,\rho} \hat{P}^{(\mu)}_{\sigma\rho}\hat{E}_{p\sigma q\rho},\;\; \mu = 0,1,2,3
	\label{eq:extended-triplet-excitation-op}
\end{eqnarray}
in the following compact form:
\begin{equation}
	\hat{H}^{(1)} = \sum_{p,q}\mathbf{\tilde{H}}^{(1)}_{pq}\cdot\mathbf{\hat{\tilde{T}}}_{pq}
	\label{eq:one-body-full-triplet}
\end{equation}

The Cartesian component representation of the Pauli-Breit Hamiltonian, as given in eqs.~\ref{eq:one-body-full},~\ref{eq:one-body-full-triplet}, carries the following interpretation:
the $Z$-components of the $\hat{H}^{(SO,1)}$ and $\hat{H}^{(SO,2)}$ operators (i.e. the zeroth-spherical tensor components) shift the non-relativistic electron energies. Spin-mixing terms are due to the $\pm1$ spherical tensor components of the spin-orbit coupling Hamiltonians (i.e. the $X$ and $Y$ Cartesian components) and cause energy level splittings\cite{Schaden2023}.

The pure two-body part of the Pauli-Breit Hamiltonian (excluding the \textit{add-on} contribution) can be expressed as:
\begin{align}
	\hat{H}^{(2)}
	&= \sum_{\mu,\mu'=0,X,Y,Z}\sum_{p,q,r,s}\sum_{\sigma,\rho,\tau,\nu}
	\Bigg[
	\delta_{\mu 0}\delta_{0\mu'}\hat{P}^{(0)}_{\sigma\rho}\hat{P}^{(0)}_{\tau\nu}
	\Bigg(
	\hat{E}_{p\sigma q \rho}\hat{E}_{r\tau s \nu}\frac{1}{2}
	\bra{pq}\hat{V}^{(ee)}+\hat{H}^{(D2)}\ket{rs}
	\nonumber\\
	&\qquad\qquad
	-\hat{E}_{q\sigma p \rho}\hat{E}_{r\tau s \nu}\frac{1}{4}
	\bra{pq}\hat{H}^{(cont)}\ket{rs}
	\Bigg)+
	\nonumber\\
	&\quad
	+(1-\delta_{0\mu'}-\delta_{\mu 0}+\delta_{\mu 0}\delta_{0\mu'})
	\hat{P}^{(\mu)}_{\sigma\rho}\hat{P}^{(\mu')}_{\tau\nu}
	\hat{E}_{p\sigma q \rho}\hat{E}_{r\tau s \nu}
	\frac{1}{2}
	\bra{pq}\hat{H}^{(SS,\mu\mu')}\ket{rs}+
	\nonumber\\
	&\quad
	+\delta_{0\mu'}(1-\delta_{\mu 0})
	\hat{P}^{(\mu)}_{\sigma\rho}\hat{P}^{(\mu')}_{\tau\nu}
	\hat{E}_{p\sigma q \rho}\hat{E}_{r\tau s \nu}
	\frac{1}{2}
	\bra{pq}\hat{H}^{(orb,2,\mu)}\ket{rs}+
	\nonumber\\
	&\quad
	+\delta_{0\mu}(1-\delta_{\mu' 0})
	\hat{P}^{(\mu)}_{\sigma\rho}\hat{P}^{(\mu')}_{\tau\nu}
	\hat{E}_{p\tau r \nu}\hat{E}_{q\sigma s \rho}
	\frac{1}{2}
	\bra{pq}\hat{H}^{(orb,2,\mu')}\ket{rs}
	\Bigg]
	\label{eq:two-body}
\end{align}

Similarly to one-body terms, a more compact form of eq.~\ref{eq:two-body} can be produced by defining the \textit{Pauli-Breit interaction matrix} as follow
\begin{eqnarray}
	\hat{\mathbf{G}} = \frac{1}{2}\begin{pmatrix}
		\hat{V}^{(ee)}+\hat{H}^{(D2)}-\frac{1}{2}\hat{H}^{(cont)} & \hat{H}^{(orb,2,x)} &  \hat{H}^{(orb,2,y)} &  \hat{H}^{(orb,2,z)} \\
		\hat{H}^{(orb,1,X)} & \hat{H}^{(SS,XX)} &  \hat{H}^{(SS,xy)} &  \hat{H}^{(SS,xz)} \\
		\hat{H}^{(orb,1,Y)}& \hat{H}^{(SS,yx)} &  \hat{H}^{(SS,yy)} &  \hat{H}^{(SS,yz)} \\
		\hat{H}^{(orb,1,Z)}& \hat{H}^{(SS,zx)} &  \hat{H}^{(SS,zy)} &  \hat{H}^{(SS,zz)} \\
	\end{pmatrix}
	\label{eq:4-pauli-matrix}
\end{eqnarray}
and the \textit{contraction matrix}:
\begin{eqnarray*}
	\mathbf{C} = \begin{pmatrix}
		\delta_{\mu 0}\delta_{0\mu'} & \delta_{0\mu}(1-\delta_{\mu' 0})&   \delta_{0\mu}(1-\delta_{\mu' 0}) &   \delta_{0\mu}(1-\delta_{\mu' 0}) \\
		\delta_{0\mu'}(1-\delta_{\mu 0}) &(1-\delta_{0\mu'}-\delta_{\mu 0}+\delta_{\mu 0}\delta_{0\mu'}) & (1-\delta_{0\mu'}-\delta_{\mu 0}+\delta_{\mu 0}\delta_{0\mu'}) &  (1-\delta_{0\mu'}-\delta_{\mu 0}+\delta_{\mu 0}\delta_{0\mu'}) \\
		\delta_{0\mu'}(1-\delta_{\mu 0})& (1-\delta_{0\mu'}-\delta_{\mu 0}+\delta_{\mu 0}\delta_{0\mu'}) & (1-\delta_{0\mu'}-\delta_{\mu 0}+\delta_{\mu 0}\delta_{0\mu'}) &  (1-\delta_{0\mu'}-\delta_{\mu 0}+\delta_{\mu 0}\delta_{0\mu'}) \\
		\delta_{0\mu'}(1-\delta_{\mu 0})& (1-\delta_{0\mu'}-\delta_{\mu 0}+\delta_{\mu 0}\delta_{0\mu'}) & (1-\delta_{0\mu'}-\delta_{\mu 0}+\delta_{\mu 0}\delta_{0\mu'}) &  (1-\delta_{0\mu'}-\delta_{\mu 0}+\delta_{\mu 0}\delta_{0\mu'}) \\
	\end{pmatrix}
	\label{eq:contraction-matrix}
\end{eqnarray*}
which yields the two-body Hamiltonian in a compact form:
\begin{equation}
	\hat{H}^{(2)} = \sum_{\mu,\mu'=0,X,Y,Z}\sum_{p,q,r,s} C_{\mu\mu'} \hat{\tilde{T}}^{(\mu)}_{pq}\hat{G}^{(\mu\mu')}_{pqrs} \hat{\tilde{T}}^{(\mu')}_{rs}
	\label{eq:two-body2}
\end{equation}
where $\hat{G}^{(\mu\mu')}_{pqrs} = \bra{pq}\hat{G}^{(\mu\mu')}\ket{rs}$. In the derivation of eq.~\ref{eq:two-body2} we used fermionic commutation relations and symmetries of the $U(2N)$ Lie group generators.
The two-body Hamiltonian can be thus written as a double-contraction of rank-6 tensor $\hat{G}$ with rank-2 contraction matrix $\mathbf{C}$ and rank-2 extended triplet excitation operators:
\begin{align}
	\hat{H}^{(2)} =  \mathbf{C} : \hat{\tilde{T}} :\hat{\mathbf{G}}  :\hat{\tilde{T}}
	\label{eq:two-body-3}
\end{align}
For PB Hamiltonian matrix element calculations CFOUR or OpenMolcas quantum chemistry software~\cite{openmolcas2023} can be used, although some parts, such as the two-body spin-orbit coupling matrix elements must be calculated elsewhere, with custom methods.

\subsection{Majorana representation of the Pauli-Breit Hamiltonian and its double-factorization}
Having expressed the PB Hamiltonian in a compact form using triplet excitation operators and $\mathcal{U}(N)\times\mathcal{U}(2)$ generators, it is convenient to introduce a further transformation to the Majorana representation, which offers practical advantages for quantum circuit design. Majorana operators are Hermitian and can be defined via the spin-orbital excitation operators as follows:
\begin{eqnarray}
	\hat{a}_{p\sigma} = \frac{1}{2}\left( \hat{\gamma}_{p\sigma 0} + \hat{\gamma}_{p\sigma 1}\right) \\
	\hat{a}^{\dagger}_{p\sigma} = \frac{1}{2}\left( \hat{\gamma}_{p\sigma 0} - \hat{\gamma}_{p\sigma 1}\right) \\	
\end{eqnarray}
with $\left[\hat{\gamma}_{p\sigma u},\hat{\gamma}_{q\rho v}\right]_+ = 2 \delta_{pq} \delta_{\sigma\rho}\delta_{uv}$.
It is sufficient to show the ladder-to-Majorana transformation for operators that mix spin components of the spin-orbital basis. For this demonstration, we choose the one-body spin-orbit coupling operator. In the Majorana representation, the respective components of the spin-orbit coupling can be written as:
\begin{eqnarray}
\hat{H}^{(SO,1,X)} = \frac{i}{4}\sum_{p,q=1}^N H^{(SO,1,X)}_{p,q}\sum_{\sigma,\rho}\hat{\gamma}_{p\sigma 0}P^{(1)}_{\sigma\rho}\hat{\gamma}_{q\rho 1} \\
\hat{H}^{(SO,1,Y)} = \frac{i}{4}\sum_{p,q=1}^N H^{(SO,1,Y)}_{p,q}\sum_{\sigma,\rho}\hat{\gamma}_{p\sigma 0}P^{(2)}_{\sigma\rho}\hat{\gamma}_{q\rho 1} + \hat{R}\\
\hat{H}^{(SO,1,Z)} = \frac{i}{4}\sum_{p,q=1}^N H^{(SO,1,Z)}_{p,q} \sum_{\sigma,\rho}\hat{\gamma}_{p\sigma 0}P^{(3)}_{\sigma\rho}\hat{\gamma}_{q\rho 1}
	\label{eq:SO-1-majorana}
\end{eqnarray}
where $\hat{R} = -\sum_{p=1}^N \sum_{\sigma\rho} \sum_{s,t} H^{(SO,1,Y)}_{p,p}\left(\hat{\gamma}_{p\sigma s}\hat{P}^{(1)}_{\sigma\rho}\hat{\gamma}_{p\rho t}\left(\mathbf{1}_{st}+\hat{X}_{st}\right)\right)$. $\hat{X}$ is the Pauli-X matrix. See Appendix~\ref{sec:AppendixA} for a detailed derivation. For many systems, it is reasonable to assume that the spin-orbit coupling constant $\xi(r)$ is independent of the electronic position, yielding an effective spin-orbit constant. Under this assumption, symmetry implies that $\hat{R}=0$: the $\hat{L}_Y$ component of the electronic angular momentum transforms according to an antisymmetric representation in the orthonormal spin-orbital basis \cite{Fedorov2003}. In the spin-adapted zeroth-order basis, labeled by irreducible representations of the molecular symmetry group, the diagonal matrix elements of the spin-orbit Hamiltonian therefore vanish. For convenience of presentation we set $\hat{R}=0$ in the following, however the $R\neq 0$ case requires only formal additional manipulations.

Ultimately, the one-body Hamiltonian can be expressed with Majorana operators as follows:
\begin{eqnarray}
		\hat{H}^{(1)} = \frac{i}{4}\sum_{p,q=1}^N \sum_{\mu = X,Y,Z,0}H^{(1,\mu)}_{p,q}  \sum_{\sigma,\rho}\hat{\gamma}_{p\sigma 0} P^{(\mu)}_{\sigma\rho}\hat{\gamma}_{q\rho 1}
		\label{eq:one-body-ham-majorana}
\end{eqnarray}
The expansion in eq.~\eqref{eq:one-body-ham-majorana} over the indices $\mu$ and $\sigma,\rho$ spans the $\mathbb{C}^2$ spin space, corresponding to an expansion in the Pauli-matrix basis, with coefficients given as $H^{(1,\mu)}_{p,q}\hat{\gamma}_{p\sigma 0}\hat{\gamma}_{q\rho 1}$. The Hermitian matrix $H^{(1,\mu)}_{p,q}$ can be spectrally decomposed as
\begin{eqnarray}
	H^{(1,\mu)}_{p,q} = \sum_{l=1}^{N}f^{(1,\mu)}_lv_{l\mu p}v^{T}_{l\mu q}
	\label{eq:h1-decomposition}
\end{eqnarray}

Alternatively, the Pauli matrix acting on the spin space in eq.~\ref{eq:one-body-ham-majorana} can be incorporated in the spectral decomposition to give:
\begin{eqnarray}
	\tilde{H}^{(1)}_{p\sigma q\rho} = \sum_{\mu = X,Y,Z,0}\ H^{(1,\mu)}_{p,q}P^{(\mu)}_{\sigma\rho} = \sum_{l=1}^{2N}\tilde{f}^{(1)}_lv_{l p\sigma}v^{T}_{l q\rho}
	\label{eq:h1-decomposition-spin}
\end{eqnarray} 
In the former case shown in eq.~\ref{eq:h1-decomposition}, the decomposed Hamiltonian has $4N$ terms in the Majorana representation, whereas the latter case yields a Hamiltonian with $2N$ terms.
Note that the contraction shown in eq.~\ref{eq:h1-decomposition-spin} presents an expansion in the $SU(2)$ basis for the spin degrees of freedom.

The purpose of spectral decomposition of Hamiltonian matrix given by eq.~\ref{eq:h1-decomposition} is to reduce the number of linear combination of unitaries (LCU) expansion terms. Having this in mind, we note that the two ways of spectral decomposition  given by eqs. ~\ref{eq:h1-decomposition},~\ref{eq:h1-decomposition-spin} can carry benefits depending on the specifics of the problem studied. For the typical  molecular Hamiltonians with spin-orbit coupling, it is advantageous to use the form given by eq.~\ref{eq:h1-decomposition}. 
The truncated Hamiltonian can be then written in the following form:
\begin{eqnarray}
	\hat{H}^{(1)} = \frac{i}{4}\sum_{l=1}^{L} \sum_{\mu = X,Y,Z,0}f_{l\mu}  P^{(\mu)}_{\sigma\rho} \vec{\hat{\gamma}}_{l\mu \sigma 0}\vec{\hat{\gamma}}_{l\mu \rho 1} + \hat{\epsilon}
	\label{eq:one-body-majorana-factorized1}
\end{eqnarray}

where $\epsilon(H^{(1)})=||\hat{\epsilon}(H^{(1)})||=	||\hat{H}^{(1)} - \frac{i}{4}\sum_{l=1}^{L} \sum_{\mu = X,Y,Z,0}f_{l\mu}  P^{(\mu)}_{\sigma\rho} \vec{\hat{\gamma}}_{l\mu \sigma 0}\vec{\hat{\gamma}}_{l\mu \rho 1}|| $ is the error controlled by the one-body Hamiltonian truncation parameter $L<N$. In eq.~\ref{eq:one-body-majorana-factorized1} we introduced Majorana vectors defined as:

\begin{eqnarray}
	\vec{\gamma}_{l\mu\sigma x} = \sum_{p=1}^N  v_{l\mu p} \hat{\gamma}_{p\sigma x}
\end{eqnarray}

We denote $\hat{\Gamma}_{l\mu\sigma\rho} := \vec{\hat{\gamma}}_{l\mu \sigma 0}\vec{\hat{\gamma}}_{l\mu \rho 1}$ to express the truncated factorized Hamiltonian given in eq.\ref{eq:one-body-majorana-factorized1} in the following form:

\begin{eqnarray}
	\hat{H}^{(1)} = \sum_{l=1}^L \sum_{\mu=0}^3 f_{l\mu} \sum_{\sigma,\rho = 0}^1P^{(\mu)}_{\sigma\rho} \hat{\Gamma}^{(l\mu)}_{\sigma\rho} 
	\label{eq:one-body-factorized-2}
\end{eqnarray}

For transforming the two-body part of the PB Hamiltonian to the Majorana representation, we denote the mapping between the triplet excitation operators and Majorana operators, written as follows:
\begin{eqnarray}
	\hat{\tilde{T}}^{(\mu)}_{p,q} = \sum_{\sigma,\rho}\sum_{x,y}t_{\sigma,\rho,x,y}^{(\mu)}\hat{\gamma}_{p\sigma x}\cdot\hat{\gamma}_{q\rho y}
	\label{triplet-to-majorana}
\end{eqnarray}

The two-body Hamiltonian given in eq.~\ref{eq:two-body2} can be expressed as 
\begin{align}
	\hat{H}^{(2)}
	&= \sum_{\mu,\mu'=0,X,Y,Z}
	\sum_{p,q,r,s}
	C_{\mu\mu'}\,
	\tilde{\hat{T}}^{(\mu)}_{pq}\,
	\hat{G}^{(\mu\mu')}_{pqrs}\,
	\tilde{\hat{T}}^{(\mu')}_{rs}
	\nonumber\\
	&= \sum_{\mu,\mu'=0,X,Y,Z}
	\sum_{p,q,r,s}
	\sum_{\sigma,\rho,\sigma',\rho'}
	\sum_{x,y,x',y'}
	C_{\mu\mu'}\,
	\hat{G}^{(\mu\mu')}_{pqrs}\,
	t_{\sigma,\rho,x,y}^{(\mu)}\,
	t_{\sigma',\rho',x',y'}^{(\mu')}\,
\hat{	\gamma}_{p\sigma x}\hat{\gamma}_{q\rho y}
	\hat{\gamma}_{r\sigma' x'}\hat{\gamma}_{s\rho' y'}
	\label{eq:two-body-majorana}
\end{align}

and further double-factorized to the following form:
\begin{equation}
	\hat{H}^{(2)} = \sum_{m=1}^M \left(\sum_{l=1}^{L(m)} \sum_{\mu,\mu'=0,X,Y,Z} f^{(m,l,\mu,\mu')}\left(\sum_{p}v_{p,\sigma,0}^{(m,l,\mu,\mu')}\hat{\gamma}_{p\sigma 0}\right)\left(\left(\sum_{q}v_{q,\rho,1}^{(m,l,\mu,\mu')}\right)^T\hat{\gamma}_{q\rho 1}\right)\right)^2 +\hat{\epsilon}\left(M,\mathbf{L}\right)
	\label{eq:two-body-DF-spin}
\end{equation}
which takes a convenient form expressed through the one-body terms as follows:
\begin{equation}
	h^{(1,m)} = \sum_{l=1}^{L(m)}  \sum_{\mu,\mu'=0,X,Y,Z} f^{(m,l,\mu,\mu')}\left(\sum_{p,\sigma }v_{p,\sigma,0}^{(m,l,\mu,\mu')}\hat{\gamma}_{p\sigma 0}\right)\left(\sum_{q,\rho}(v_{q,\rho,1}^{(m,l,\mu,\mu')})^T\gamma_{q\rho 1}\right) 
\end{equation} 

and denoting: $\sum_{p}v_{p,\sigma,0}^{(m,l,\mu,\mu')}\gamma_{p\sigma 0} :=\vec{\gamma}_{m,l,\sigma,\mu,\mu',0}$ and $\hat{\Gamma}^{(m,l,\mu,\mu')}_{\sigma\rho} := \vec{\gamma}_{m,l,\mu,\mu'\sigma, 0}\vec{\gamma}_{m,l,\mu,\mu'\rho, 1}$, we can express the one-body components  as:
\begin{eqnarray}
	\hat{h}^{(1,m)} = \frac{i}{4}\sum_{l=0}^{L(m)}  \sum_{\mu,\mu'=0,X,Y,Z} f^{(m,l,\mu,\mu')}\sum_{\sigma \rho} \hat{\Gamma}^{(m,l,\mu,\mu')}_{\sigma\rho}
\end{eqnarray}
and the  two-body Hamiltonian in a compact form:
\begin{eqnarray}
	\hat{H}^{(2)} = \sum_{m=1}^M \left(\sum_{l=1}^{L(m)} \sum_{\mu,\mu'=0,X,Y,Z} f^{(m,l,\mu,\mu')}\sum_{\sigma\rho }\hat{\Gamma}^{(m,l,\mu,\mu')}_{\sigma\rho}\right)^2 +\hat{\epsilon}\left(M,\mathbf{L}\right)
	\label{eq:two-body-DF-spin-2}
\end{eqnarray}
further written as a linear combination of  Chebyschev polynomials of the respective one-body operators:
\begin{eqnarray}
	\hat{H}^{(2)} = \sum_{m=1}^{M} ||h^{(1,m)}||^2 T_2\left(\frac{\hat{h}^{(1,m)}}{||h^{(1,m)}||}\right)
	\label{eq:two-body-chebyschev-qubitization2}
\end{eqnarray}
where $T_2(x) = 2x^2-1$ is the Chebyschev polynomial of the first kind and $\hat{\epsilon}\left(M,\mathbf{L}\right)$ denotes truncation error.
The free terms (those proportional to identity) appearing upon transition from eq.~\ref{eq:two-body-DF-spin-2} to eq.~\ref{eq:two-body-chebyschev-qubitization2} can be discarded in the simulation, as long as we keep their values in classical memory.

\section{Qubit mapping}
The spin-orbital creation and annihilation operators $\lbrace \hat{a}_{p\sigma}\rbrace_{p=1,2,...,N; \sigma=\pm \frac{1}{2}}$ are mapped to Pauli operators via the \textit{Jordan-Wigner} encoding. In doing so we propose to order the spin-orbital basis taking the spin index as major in \textit{orbital-major} (\textit{OM}) ordering. This ordering carries advantages over the \textit{spin-major} (\textit{SM}) encoding used so far in the literature, as discussed further. 
The ordered spin-orbital basis in the \textit{OM} ordering is given as:
\begin{equation}
	\mathcal{B}_{OM}=\lbrace \ket{1\alpha},  \ket{1\beta},  \ket{2\alpha},  \ket{2\beta}, ...,  \ket{N\alpha}, \ket{N\beta}\rbrace
\end{equation}
The ordered spin-orbital basis in the SM ordering is given as:
\begin{eqnarray}
	\mathcal{B}_{SM}=\lbrace \ket{1\alpha},  \ket{2\alpha},...\ket{N\alpha},  \ket{1\beta},  \ket{2\beta},..., \ket{N\beta}\rbrace
\end{eqnarray}
Qubit encoding for the respective basis orderings can be written as
\begin{eqnarray}
	\hat{a}_{p\sigma} = \hat{a}_{\sigma N+p} = \otimes_{j=0}^{\sigma N + p -1} \hat{Z}_j \otimes \hat{\sigma}^+_{\sigma N + p}\otimes \hat{\mathbf{1}}_{\sigma N+p+1} \otimes \hdots \otimes  \hat{\mathbf{1}}_{2N}, \; SM\\
	\hat{a}_{p\sigma} = \hat{a}_{2(p-1)+\sigma} = \otimes_{j=0}^{2(p-1)+\sigma-1} \hat{Z}_j \otimes \hat{\sigma}^+_{2(p-1)+\sigma}\otimes \hat{\mathbf{1}}_{2(p-1)+\sigma+1} \otimes \hdots \otimes  \hat{\mathbf{1}}_{2N}, \; OM\\
	\label{eq:JW-encoding}
\end{eqnarray} 
Majorana operators can be now expressed as Pauli strings. We first use the following identity:
\begin{eqnarray}
	\hat{\gamma}_{p\sigma x}\hat{\gamma}_{q\rho y} = \left(-i\right)^{x+y}\left[\hat{a}_{p\sigma}\hat{a}_{q\rho}+(-1)^y\hat{a}_{p\sigma}\hat{a}^{\dagger}_{q\rho}+(-1)^x\hat{a}^{\dagger}_{p\sigma}\hat{a}_{q\rho}+(-1)^{(x+y)}\hat{a}^{\dagger}_{p\sigma}\hat{a}^{\dagger}_{q\rho}\right]
	\label{eq:majorana-ladder}
\end{eqnarray}
In the case of the \textit{SM} ordering we split the encoded operator into the following cases:
\scriptsize
\begin{equation}
	\hat{\gamma}_{p\sigma x}\hat{\gamma}_{q\rho y}
	=
	\begin{cases}
		(-i)^{x+y}\,
		\hat{\mathbf{1}}_{[1,p-1]}
		\otimes \hat{C}^{(1)}_p(x)
		\otimes \hat{Z}_{[p+1,N+q-1]}
		\otimes \hat{C}^{(2)}_{N+q}(y)
		\otimes \hat{\mathbf{1}}_{[N+q+1,2N]},
		& \sigma < \rho, \\[0.6em]
		(-i)^{x+y}\,
		\hat{\mathbf{1}}_{[1,N\sigma+p-1]}
		\otimes \hat{C}^{(1)}_{N\sigma+p}(x)
		\otimes \hat{Z}_{[N\sigma+p+1,N\sigma+q-1]}
		\otimes \hat{C}^{(2)}_{N\sigma+q}(y)
		\otimes \hat{\mathbf{1}}_{[N\sigma+q+1,2N]},
		& \sigma = \rho,\; p \ge q .
	\end{cases}
	\label{eq:majorana-SM}
\end{equation}
\normalsize
where
\begin{eqnarray}
	\hat{C}^{(1)}_p(x) := \hat{X}_p \cdot \frac{(-1)^x-1}{2}-i\hat{Y}_p\cdot \frac{1+(-1)^x}{2}\\
	\hat{C}^{(2)}_{N+q}(y) := \hat{X}_{N+q} \cdot \frac{1+(-1)^y}{2}-i\hat{Y}_{N+q}\cdot \frac{1-(-1)^y}{2}\\
	\label{eq:majorana-encoding-C}
\end{eqnarray}
and the bracket notation for operators, e.g. $\hat{\mathbf{1}}_{[1,N\sigma+p-1]}$ denotes their respective qubit range (closed). In the derivation of eq.~\ref{eq:majorana-SM} we used the following relations between Pauli operators:
\begin{eqnarray}
	\hat{Z}\hat{P}^{\pm} =\pm\hat{P}^{\pm}\\
	\left[\hat{Z},\hat{P}^{\pm}\right]_- =\pm 2\hat{P}^{\pm}\\
	\left[\hat{Z},\hat{P}^{\pm}\right]_+ =0\\
	\hat{P}^{(j)} \hat{P}^{(k)} = i\epsilon_{jkl}\hat{P}_l, \qquad j,k,l = x,y,z\\
\end{eqnarray}
where $\hat{P}^{(x)} \equiv \hat{X}$ etc. 
For the new \textit{OM} ordering the product of Majorana operators has the following form:

	\begin{align}
	\hat{\gamma}_{p\sigma x}\hat{\gamma}_{q\rho y} =\left(-i\right)^{x+y} \hat{\mathbf{1}}_{[1,2(p-1)+\sigma-1]}\otimes \hat{C}^{(1)}_{2(p-1)+\sigma}(x) \otimes \\ \otimes\hat{Z}_{[2(p-1)+\sigma+1,2(q-1)+\rho-1]}\otimes \hat{C}^{(2)}_{2(q-1)+\rho}(y)  \otimes \hat{\mathbf{1}}_{[2(q-1)+\rho+1,2N]}, \quad p \leq q; \\
		\label{eq:majorana-OM}
	\end{align}
such that, for example when $x=0,y=1$, $p=q$ and $\sigma < \rho$ we have:
\begin{eqnarray}
		\hat{\gamma}_{p\sigma 0}\hat{\gamma}_{p\rho 1} = i\hat{Y}_{2(p-1)}\otimes \hat{Y}_{2(p-1)+1}
\end{eqnarray}
in a notation where identity operators have been dropped.
Comparing the \textit{OM} with \textit{SM} encoding we notice that the number of non-identity Pauli operators in a given encoded Majorana operator product is different and given by the formulas:

\begin{eqnarray}
	n_P(SM) = (1-\delta_{\sigma\rho})N +|{p-q}|, \qquad SM \\
	n_P(OM) = 2|p-q|+|\rho-\sigma|, \qquad OM
	\label{eq:compare-OM-SM}
\end{eqnarray}
where $\hat{C}^{(1/2)}$ are counted as one Pauli operator. The \textit{Pauli weight} difference between the two encoding is given thus by:
\begin{eqnarray}
	\Delta n_C = |n_P(SM)-n_P(OM)| = |(N-1)| \cdot |\sigma-\rho| -|q-p|
	\label{eq:OM-SM-diff}
	\end{eqnarray}
which is largest when $q=p$ and smallest when $q=N$ and $p=0$. For this reason we consider \textit{OM} when $|q-p|$ is small and when spin indices are different. This situation corresponds to a physical system in which the off-diagonal spin-orbit coupling is weak. On the other hand, the \textit{SM} ordering is better when the off-diagonal spin-orbit operator mixes significantly orbitals with distant indices. Because molecular or atomic orbitals are usually ordered in ascending energy, the strength of their coupling is proprortional to orbital energy difference. In such a situation the \textit{OM} ordering is advisable. Normally the Clifford gate cost is neglected in Fault-Tolerant resource estimation. We however are mindful of this particular Clifford cost, as it becomes relevant in other parts of the circuit.  When time-dependent interaction with an external electromagnetic field is considered, one may switch between the \textit{OM} and \textit{SM} orderings as the simulation progresses, guided by the orbital coupling strength.

\section{Block encoding the double-factorized Pauli-Breit Hamiltonian}
\label{sec:encoding}
\begin{figure}[h] 
	\centering
	\includegraphics[width=0.6\linewidth]{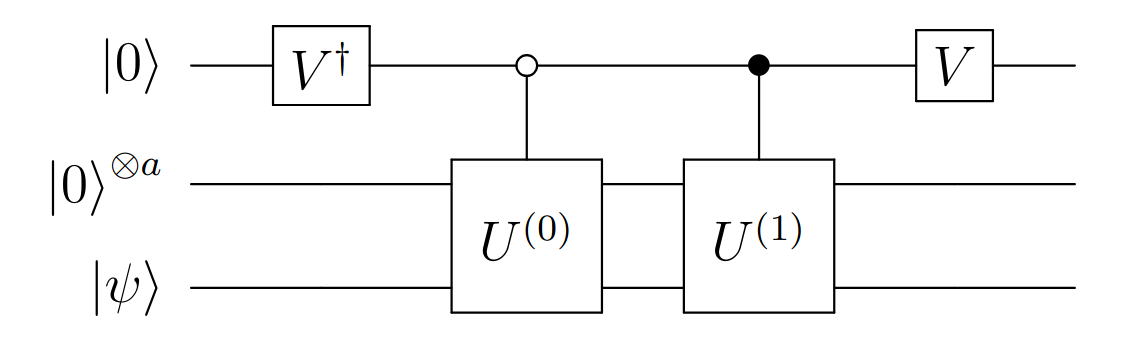} 
	\caption{Addition of block-encoding of  $\hat{H}^{(1)}$ represented by $U^{(0)}$ and   $\hat{H}^{(2)}$ represented by $U^{(1)}$. Here $V$ is the Hadamard gate.}
	\label{fig:Addition}
\end{figure}

The PB Hamiltonian can be written as a sum of one body terms $\hat{H}^{(1)}$ defined in eq.~\ref{eq:one-body-factorized-2} and two-body terms $\hat{H}^{(2)}$ defined in eq.~\ref{eq:two-body-chebyschev-qubitization2}:

\begin{equation}
	\hat{H} = \hat{H}^{(1)}+\hat{H}^{(2)}
	\label{eq:pb}
\end{equation}
In this section we construct a unitary circuit $\mathbb{B}[\hat{H}]$ block-encoding the PB Hamiltonian given in eq.\ref{eq:pb}. 
In doing so we block-encode $\hat{H}^{(2)}$ and $\hat{H}^{(1)}$ separately and combine them through addition of block-encoding unitaries, which introduces one ancilla qubit, as shown in Fig.\ref{fig:Addition}. The two-body  Hamiltonian is block-encoded in unitary circuit $U^{(1)}$ with qubitization~
\cite{low2017}, using block-encoding of the respective one-body operators.

Block-encodings of one-body operators are constructed via LCU~\cite{berry2019} can be written in a general form:
\begin{eqnarray}
	\mathbb{B}[\hat{H}] = \hat{U}^{\dag}\hat{V} \hat{U}
	\label{eq:LCU}
\end{eqnarray}
where  $\hat{U}$ is the state preparation unitary and $\hat{V}$ is the selection unitary. We discuss the  construction of the respective block-encoding components of eq.~\ref{eq:LCU} in the following section. 

\subsection{State preparation}

We first construct a block-encoding of the one-body Hamiltonian. In the LCU framework, $\hat{H}^{(1)}$ is written in a double-factorized form as
\begin{eqnarray}
	\hat{H}^{(1)} = \sum_{l=1}^L \sum_{\mu=0}^3 f_{l\mu} \sum_{\sigma,\rho = 0}^1P^{(\mu)}_{\sigma\rho} \hat{\Gamma}^{(l\mu)}_{\sigma\rho} 
	\label{eq:one-body-factorized-a}
\end{eqnarray}
where $\hat{\Gamma}^{(l\mu)}_{\sigma\rho}$ is unitary. To implement the state-preparation unitary $U$ introduced in Eq.~\ref{eq:LCU}, we introduce additional quantum registers that index the coefficients $f_{l\mu}$ and $P^{(\mu)}_{\sigma\rho}$.
The composite state-preparation unitaries are defined as
\begin{eqnarray}
	\ket{f}=U_f \ket{0}_{n_L+2} = \sum_{\mu = X,Y,Z,0}\sum_{l=1}^L \sqrt{f_{l\mu}}\ket{l\mu},
	\label{eq:state-prep-f}
\end{eqnarray}
and
\begin{eqnarray}
	\ket{P}=U_p \ket{0}_{4} = \sum_{\mu = X,Y,Z,0}\sum_{\sigma,\rho=0}^1 \sqrt{P^{(\mu)}_{\sigma\rho}}\ket{\sigma\rho\mu}.
	\label{eq:state-prep-s}
\end{eqnarray}
Here $\mu$ is encoded using two qubits $\ket{x_1x_2}$, and $n_L$ denotes the number of qubits required to represent indices in the range $[0,L-1]$. We assume a $b$-bit representation of the eigenvalues $f$.
We then construct
\begin{eqnarray}
	U = U_L\left(U_f \otimes U_p\right),
\end{eqnarray}
where the auxiliary logic unitary $U_L$ enforces the constraint $\mu=\mu'$, yielding the joint state
\begin{eqnarray}
	U\ket{0} =	\sum_{\mu = X,Y,Z,0}\sum_{\sigma \rho}\sum_{l=1}^L\sqrt{f_{l\mu}P^{(\mu)}_{\sigma\rho}}\ket{l \mu \sigma \rho}.
	\label{eq:prepstate}
\end{eqnarray}
The cost of $U_L$ is independent of the problem size, i.e., $\mathcal{O}(1)$, and is therefore negligible in the overall resource accounting. Alternatively, one may synthesize the state in eq.~\ref{eq:prepstate} directly, without explicitly separating the preparations of $\ket{f}$ and $\ket{P}$.

The state in eq.~\ref{eq:state-prep-f} can be prepared using standard techniques~\cite{Gosset:2024,low2024,babbush2018} with Clifford+T cost scaling as $\mathcal{O}(L)$ and $\mathcal{O}(\sqrt{b})$ qubits. $L$ typically scales linearly with the number of spin-orbitals, $\mathcal{O}(N)$. Depending on the chosen QROM implementation, the preparation requires $\mathcal{O}(\sqrt{Lb})$ Clifford+T gates when additional $\mathcal{O}(\sqrt{Lb})$ ancilla qubits are available.

\begin{figure}[h] 
	\centering
	\includegraphics[width=0.6\linewidth]{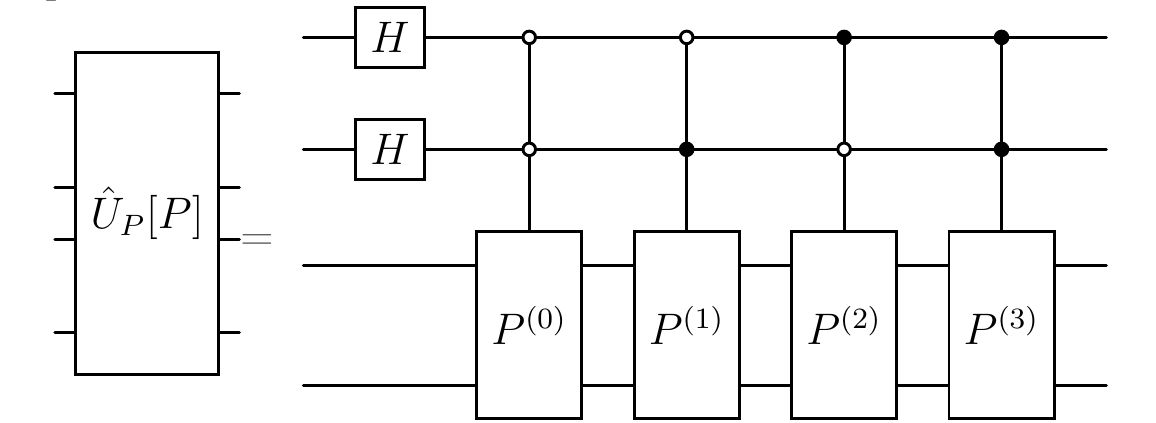}
	\caption{Quantum circuit preparing a two-qubit state with amplitudes given by the corresponding square roots of Pauli matrix elements $\sqrt{P^{(\mu)}_{\sigma\rho}}$, $\mu=0,X,Y,Z$.}
	\label{fig:Pauli_PREP}
\end{figure}

\begin{figure}[!h] 
	\centering
	\includegraphics[width=0.9\linewidth]{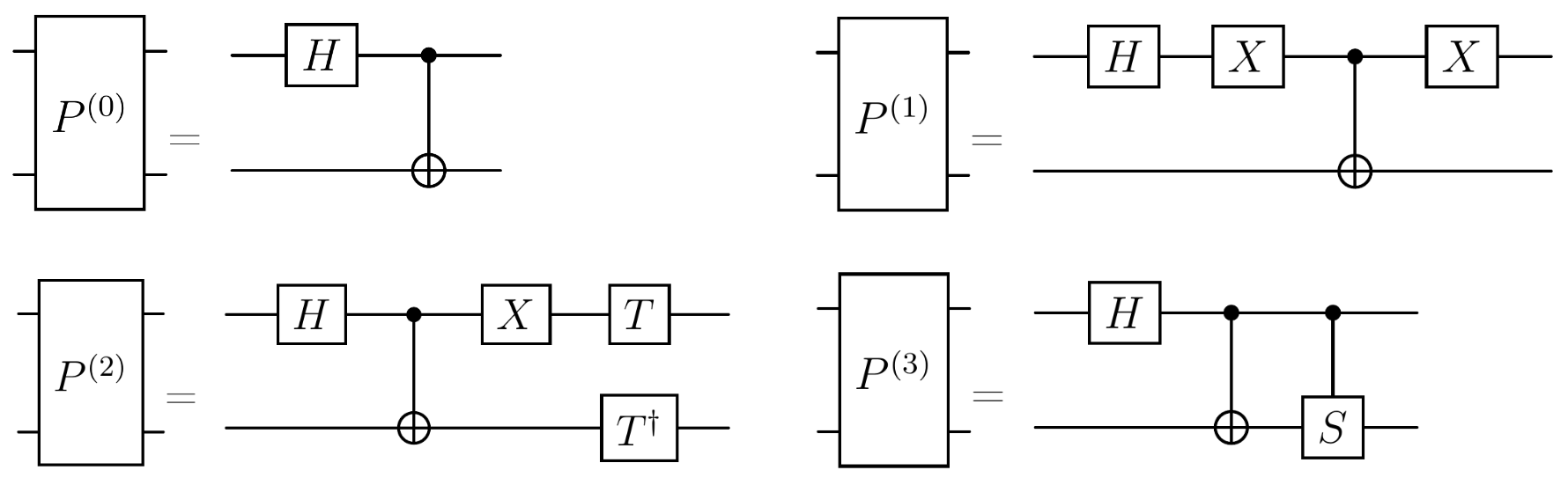}
	\caption{Circuit primitives preparing two-qubit states with amplitudes given by the corresponding elements of the Pauli matrices $\mu=0,X,Y,Z$.}
	\label{fig:Pauli}
\end{figure}

The unitary $U_p$ in eq.~\ref{eq:state-prep-s} can be implemented using the circuit shown in Fig.~\ref{fig:Pauli_PREP}.
The elementary operations $P^{(0)},P^{(1)},P^{(2)},P^{(3)}$ that generate the square roots of the Pauli-matrix elements $\sqrt{P^{(\mu)}_{\sigma\rho}}$ in Fig.~\ref{fig:Pauli_PREP} are given in Fig.~\ref{fig:Pauli}. These primitives require $\mathcal{O}(1)$ Clifford+T gates and two qubits.

\subsection{Selection of unitaries}

\subsubsection{Relativistic extension of the doubly-factorised non-relativistic electronic Hamiltonian through Pauli-SWAP networks}
\begin{figure}[!h] 
	\centering
	\includegraphics[width=0.9\linewidth]{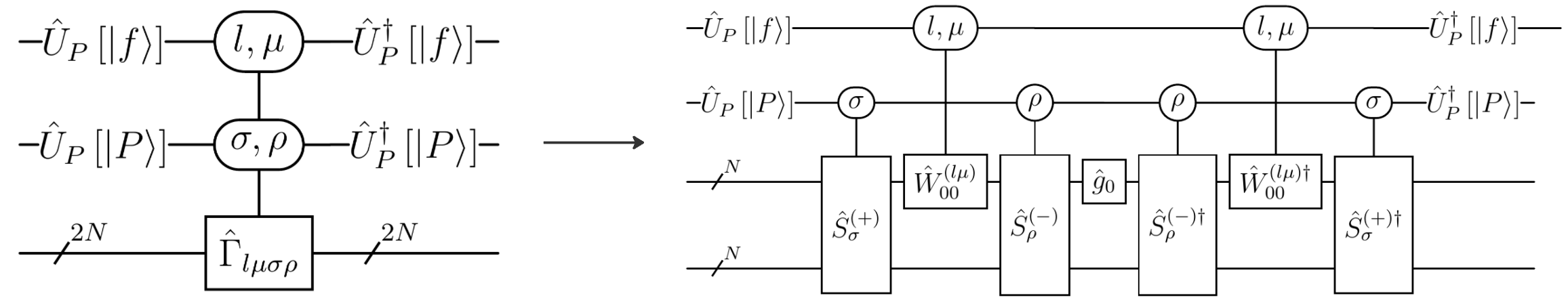}
	\caption{a) Quantum circuit directly block encoding the factorized Hamiltonian given in eq.~\ref{eq:one-body-majorana-factorized1}; 
		b) Block-encoding circuit for the same Hamiltonian employing spin-swapping networks implemented via $\hat{S}^{(\pm)}_{\sigma}$.}
	\label{fig:BE_DF}
\end{figure}

Combining the state-preparation unitary defined in eq.~\ref{eq:prepstate} with the standard SELECT unitary associated with $\hat{\Gamma}^{(l\mu)}_{\sigma\rho}$ in eq.~\ref{eq:one-body-factorized-a} yields the block-encoding circuit shown in Fig.~\ref{fig:BE_DF}a. 
Instead, we propose an alternative quantum circuit that decouples the mixed-spin control from the orbital control, as illustrated in Fig.~\ref{fig:BE_DF}b. This construction reduces the cost of the generic circuit by approximately a factor of two by explicitly accounting for the spin-dependent structure of the PB Hamiltonian. In particular, spin-dependent contributions are encoded in the Pauli matrix basis and implemented directly as qubit-level operations. This is possible because the Pauli matrices form a complete basis for operators acting on qubit Hilbert spaces. 

The specific structure of the Hamiltonian in eq.~\ref{eq:one-body-factorized-a} can therefore be exploited in conjunction with controlled-SWAP operations, which, depending on the spin indices, encode the appropriate Pauli-matrix elements and apply the corresponding spin-independent orbital operations for each term in the linear combination of unitaries.

Our proposed construction proceeds as follows. The selection unitary can be written in the general form
\begin{eqnarray}
	\hat{V} = \sum_{l=1}^L \sum_{\mu=0}^3 \sum_{\sigma\rho=0}^1 
	|l\mu \sigma \rho\rangle \langle l\mu \sigma \rho| \otimes \hat{\Gamma}_{\sigma\rho}^{(l\mu)} ,
	\label{eq:US}
\end{eqnarray}

Here $\hat{\Gamma}^{(l\mu)}_{\sigma\rho} = \vec{\gamma}_{l,\mu,\sigma,0}\vec{\gamma}_{l,\mu,\rho,1}$, and the transformed Majorana operators are defined as
$\hat{\gamma}_{l\mu\sigma x} = \sum_{p=1}^N v_p^{(l\mu)} \hat{\gamma}_{p\sigma x}$.
For the proposed construction to be valid, it is required that $\hat{\Gamma}^{(l\mu)}_{\sigma\rho}$ is unitary. Although $\hat{\Gamma}^{(l\mu)}_{\sigma\rho}$ is expressed as a product of linear combinations of Majorana operators (Majorana vectors), the properties of the Majorana representation enables the identification of the linear combinations for which $\hat{\Gamma}^{(l\mu)}_{\sigma\rho}$ is unitary.
Specifically, the Majorana vector admits a sum-to-product transformation of the form
\begin{equation}
	\hat{\gamma}_{l\mu\sigma x} = \sum_{p=1}^N v_p^{(l\mu)}\hat{\gamma}_{p\sigma x} 
	= \tilde{U}^{(l\mu)}_{\sigma,x} \hat{g}_x \tilde{U}^{(l\mu)\dagger}_{\sigma,x} ,
	\label{eq:gamma-u} 
\end{equation}
where $\hat{\gamma}_{l\mu\sigma x}^2 = \mathbf{1}$ is hermitian-unitary as a consequence of the defining properties of Majorana operators and since $v_p^{(l\mu)}$ are elements of unit length vectors. $\hat{g}_x$ is the spin seed operator, depending on the choice of coupling different Majorana operators, as discussed below.

Each unitary $\tilde{U}^{(l\mu)}_{\sigma,x}$ has a decomposition into local operations, the structure of which is determined by the chosen qubit encoding and the Majorana-operator coupling scheme,
\begin{equation}
	\tilde{U}^{(l\mu)}_{\sigma,x} = \prod_{p=1}^{N-1}\hat{V}^{(p)}_{\sigma,x}\left(\theta_{l\mu p}\right).
	\label{eq:u-v}
\end{equation}
Each operator $\hat{V}^{(l\mu)}_{k,x}$ is generated by a bilinear Majorana term along a prescribed Majorana chain,
\begin{eqnarray}
	\hat{V}^{(l\mu)}_{k,x}
	= \exp\left(\theta_{l\mu k}\hat{\gamma}_{p(k)\sigma(k) x}\hat{\gamma}_{p(k+1)\sigma(k+1) x}\right),
	\label{eq:u-anzatz-gamma}
\end{eqnarray}
where the mapping functions $p(k)$ and $\sigma(k)$ are defined the equations given belo
\begin{eqnarray}
	p(0) = 0,\\
	p(k>0)=\lfloor\frac{k}{4}\rfloor +1,\\
	\sigma(k) = 0,\; \left(k/2\right)\mod 2 = 0\;:\alpha,\\
	\sigma(k) = 1,\; \left(k/2\right)\mod 2 = 1\;:\beta.
	\label{eq:mapping-functions}
\end{eqnarray}
This choice generates the following chains of Majorana operators in the\textit{ OM } qubit encoding,
\begin{eqnarray}
	\hat{\gamma}_{p\sigma x}\hat{\gamma}_{p+1\alpha x}
	\rightarrow \hat{\gamma}_{p+1\alpha x}\hat{\gamma}_{p+1\beta x}
	\rightarrow \hat{\gamma}_{p+1\beta x}\hat{\gamma}_{p+2\alpha x}
	\rightarrow \hat{\gamma}_{p+2\alpha x}\hat{\gamma}_{p+2\beta x}
	\rightarrow \ldots
\end{eqnarray}

The \textit{OM} encoding yields the following Majorana-to-Pauli operator mappings,
\begin{eqnarray}
	\gamma_{p\alpha 0}\gamma_{p+1\alpha 1}
	= \hat{C}^{(1)}_{2p-2}(0)\otimes \hat{Z}_{2p-1} \otimes \hat{C}^{(2)}_{2p}(1), \\
	\gamma_{p+1\alpha 0}\gamma_{p+1\beta 1}
	= \hat{C}^{(1)}_{2p}(0)\otimes \hat{C}^{(2)}_{2p+1}(1), \\
	\gamma_{p+1\beta 0}\gamma_{p+2\alpha 0}
	= \hat{C}^{(1)}_{2p+1}(0)\otimes \hat{C}^{(2)}_{2p+2}(1),
	\label{eq:majorana-to-pauli-OM}
\end{eqnarray}
which give
\begin{eqnarray}
	\hat{V}^{(l)}_{k,x}
	= \hat{C}_{p(k)}(x)\otimes \exp\left(\theta_{lk} \hat{Z}_{p(k)}\right)
	\otimes \hat{C}^{(\dag)}_{p(k)+1}(x).
\end{eqnarray}
For comparison, the SM encoding produces the generator chain
\begin{eqnarray}
	\hat{\gamma}_{p\alpha x}\hat{\gamma}_{p+1\alpha x}
	\rightarrow \hat{\gamma}_{p+1\alpha x}\hat{\gamma}_{p+2\alpha x}
	\rightarrow \hat{\gamma}_{p+2\alpha x}\hat{\gamma}_{p+3\alpha x}
	\rightarrow \hat{\gamma}_{p+3\alpha x}\hat{\gamma}_{p+4\alpha x}
	\rightarrow \ldots
\end{eqnarray}
with the corresponding Majorana-to-Pauli mappings
\begin{eqnarray}
	\gamma_{p\alpha 0}\gamma_{p+1\alpha 1}
	= \hat{C}^{(1)}_{2p-2}(0)\otimes \hat{Z}_{2p-1} \otimes \hat{C}^{(2)}_{2p}(1), \\
	\gamma_{p+1\alpha 0}\gamma_{p+1\beta 1}
	= \hat{C}^{(1)}_{2p}(0)\otimes \hat{C}^{(2)}_{2p+1}(1), \\
	\gamma_{p+1\beta 0}\gamma_{p+2\alpha 0}
	= \hat{C}^{(1)}_{2p+1}(0)\otimes \hat{C}^{(2)}_{2p+2}(1),
	\label{eq:majorana-to-pauli-SM}
\end{eqnarray}

Finally, when the \textit{OM} encoding is used, it is possible to completely avoid weight-three Pauli strings in eq.~\ref{eq:majorana-to-pauli-OM} by choosing the spin seed $\sigma(0) = \beta$, which fixes the initial condition for the construction of the Majorana-operator chain. In this case, each operator $\hat{V}^{(l)}_{k,x}$ can be decomposed exclusively into single-qubit arbitrary $Z$ rotations and Clifford operations, as in Eqs.~\ref{eq:majorana-encoding-C} and \ref{eq:majorana-OM}.

Regardless of the chosen encoding (\textit{OM} or \textit{SM}) or the specific choice of Majorana-chain construction, each local operation appearing in the product in eq.~\ref{eq:u-v} can be expressed as a Givens-type rotation. Such a rotation is implemented as an arbitrary-angle $Z$ rotation $\exp\left( \tilde{\theta}_{l\mu p}\hat{Z}_{N\sigma+p}\right)$ conjugated by two Clifford gates $\hat{C}^{(1)}_{p\sigma x}$ and $\hat{C}^{(2)}_{p\sigma x}$, yielding
$
\hat{V}^{(p)}_{\sigma,x}\left(\theta_{l\mu p}\right)
= \exp\left( \theta_{l\mu p} \hat{\gamma}_{p\sigma x}\hat{\gamma}_{p+1\sigma x}\right)
= \hat{C}_{p\sigma x}\exp\left( \tilde{\theta}_{l\mu p}\hat{Z}_{N\sigma+p}\right)\hat{C}^{(2)}_{p\sigma x}.
$

\begin{figure}[!h] 
	\centering
	\includegraphics[width=0.5\linewidth]{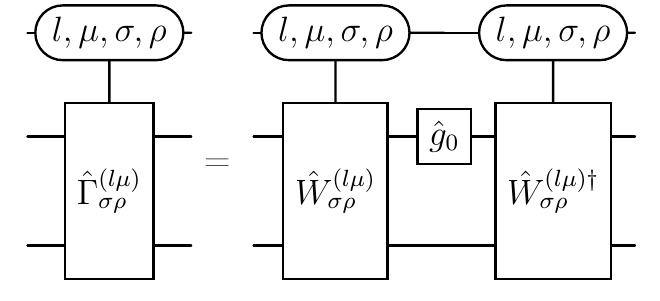}
	\caption{Quantum circuit implementing the SELECT operation.}
	\label{fig:SELECT}
\end{figure}

Using this decomposition, the operator $\hat{\Gamma}^{(l\mu)}_{\sigma\rho}$ can be written as
\begin{equation}
	\hat{\Gamma}^{(l\mu)}_{\sigma\rho} = \hat{W}^{(l\mu)}_{\sigma\rho} \hat{g}_0 \hat{W}^{(l\mu)\dagger}_{\sigma\rho},
	\label{eq:gamma-W} 
\end{equation}
with
\begin{equation}
	\hat{W}^{(l\mu)}_{\sigma\rho} = \tilde{U}^{(l\mu)}_{\sigma,0} \tilde{U}^{(l\mu)}_{\rho,1}.
	\label{eq:W-U} 
\end{equation}
Here $\hat{g}_0$ denotes the Clifford spin-seed operator constructed from the Majorana-chain seed operators introduced in eq.~\ref{eq:gamma-u}.
The resulting general SELECT circuit is shown in Fig.~\ref{fig:SELECT}.
We next seek a transformation that introduces arbitrary spin indices into the Givens rotations for all orbital indices simultaneously,
\begin{equation}
	\hat{W}^{(l\mu)}_{00}\rightarrow \hat{W}^{(l\mu)}_{\sigma\rho},
\end{equation}
such that
\begin{eqnarray}
	\tilde{U}^{(l\mu)}_{\sigma,0} \tilde{U}^{(l\mu)}_{\rho,1}
	= \hat{S}^{(+)}_{\sigma} \tilde{U}^{(l\mu)}_{0,0} \tilde{U}^{(l\mu)}_{0,1} \hat{S}^{(-)}_{\rho}.
\end{eqnarray}
Equivalently, this corresponds to the transformation
\begin{eqnarray}
	\hat{S}_{\sigma\rho}:
	\prod_{p=1}^{N-1} \hat{V}^{(p)}_{0,0}\left(\theta_{l\mu p}\right)
	\hat{V}^{(p)}_{0,1}\left(\theta_{l\mu p}\right)
	\rightarrow
	\prod_{p=1}^{N-1} \hat{V}^{(p)}_{\sigma,0}\left(\theta_{l\mu p}\right)
	\hat{V}^{(p)}_{\rho,1}\left(\theta_{l\mu p}\right).
\end{eqnarray}

\begin{figure}[!h] 
	\centering
	\includegraphics[width=0.5\linewidth]{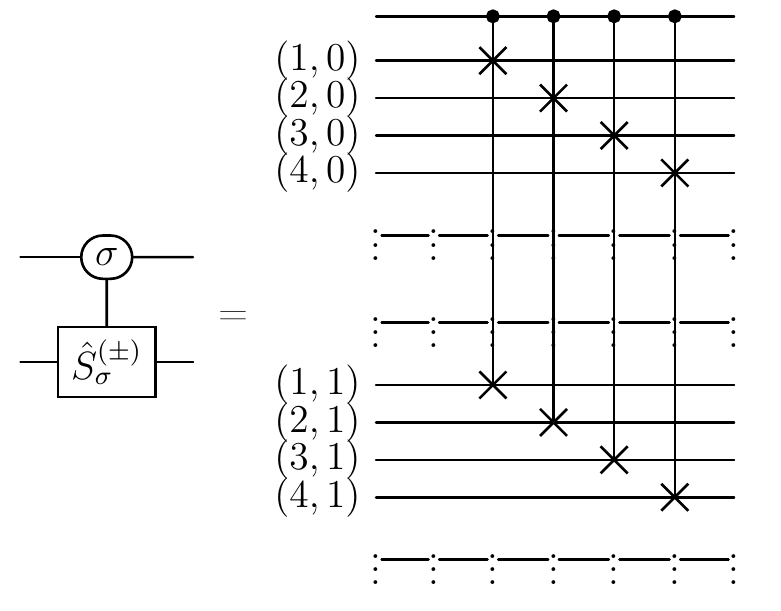}
	\caption{Quantum circuit implementing the SPIN-SWAP operations $	\hat{S}^{(\pm)}_{\sigma}$.}
	\label{fig:spin-swap}
\end{figure}

This transformation can be implemented the following primitives:
\begin{eqnarray}
	\hat{S}^{(\pm)}_{\sigma}
	= |1\rangle\langle 1|\otimes \prod_{p=1}^N SWAP[p,p+N]
	+ |0\rangle\langle 0|\otimes \mathbf{1}_{2N},
\end{eqnarray}
These SPIN-SWAP networks appear in the circuit as spin-controlled operations, as shown in Fig.~\ref{fig:BE_DF}b, which can be realized with $\mathcal{O}(1)$ T-depth and no ancilla qubits by utizing controlled-SWAP implementations discussed in Ref.~\cite{low2024}, as shown in Fig.~\ref{fig:spin-swap}.

\begin{figure}[!h] 
	\centering
	\includegraphics[width=\linewidth]{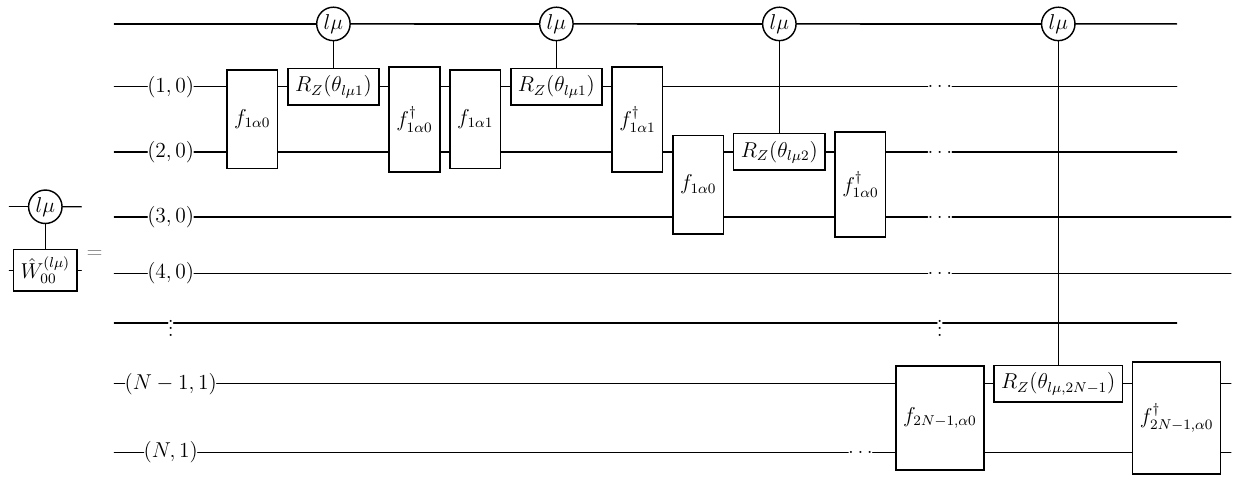}
	\caption{Quantum circuit implementing the spin-independent controlled-$\hat{W}$ operation in the OM qubit mapping. Here $f$ denote Clifford operations discussed in the main text and $R_Z(\theta)$ are $Z$-rotations}
	\label{fig:W00}
\end{figure}

Combining all components, the decomposition of $	\hat{\Gamma}^{(l\mu)}_{\sigma\rho}$ takes the form
\begin{eqnarray}
	\hat{\Gamma}^{(l\mu)}_{\sigma\rho}
	= \hat{S}^{(+)}_{\sigma}\tilde{U}^{(l\mu)}_{0,0}\tilde{U}^{(l\mu)}_{0,1}
	\hat{S}^{(-)}_{\rho}\hat{g}_0
	\hat{S}^{(-)\dagger}_{\rho}\tilde{U}^{(l\mu)\dagger}_{0,1}
	\tilde{U}^{(l\mu)\dagger}_{0,0}\hat{S}^{(+)\dagger}_{\sigma}.
\end{eqnarray}
The complete circuit is shown in Fig.~\ref{fig:BE_DF}, while a detailed decomposition of the spin-independent controlled-$\hat{W}^{(l\mu)}_{00}$ operation, 	$\hat{CW}_{00}^{(l\mu)} = \sum_{l=1}^{N}\sum_{\mu = X,Y,Z,0} \ket{l\mu}\bra{l\mu} \otimes \hat{W}^{(l\mu)}_{00}$, is illustrated in Fig.~\ref{fig:W00}.
Note that $	\hat{W}^{(l\mu)}_{00}$ factorizes into spin-independent Givens rotations:
\begin{eqnarray}
	\hat{W}^{(l\mu)}_{00}
	= \prod_{k=1}^{2N-1}\left(\hat{V}^{(l\mu)}_{k,0}\right)
	\prod_{k'=1}^{2N-1}\left(\hat{V}^{(l\mu)}_{k',1}\right).
\end{eqnarray}

which take the form: \begin{eqnarray}
	\hat{V}^{(l)}_{k,x}
	= \hat{C}_{p(k)}(x)\cdot R_Z\left(\theta_{l\mu p(k)}\right)
	\cdot \hat{C}^{(\dag)}_{p(k)+1}(x).
\label{eq:givens2}
\end{eqnarray}
as displayed in Fig.~\ref{fig:W00}. The array of angles $\theta_{l\mu k}$ shown in Eq.~\ref{eq:givens2} has dimensions $L \times 4 \times (N-1)$, where $L$ denotes the number of eigenvectors retained in the factorized representation of the one-body Hamiltonian. Each operator $\hat{V}^{(l\mu)}_{k,x}$ can be restricted to act on two qubits through an appropriate choice of generators~\cite{Arrazola2022}.

\begin{figure}[!h] 
	\centering
	\includegraphics[width=0.9\linewidth]{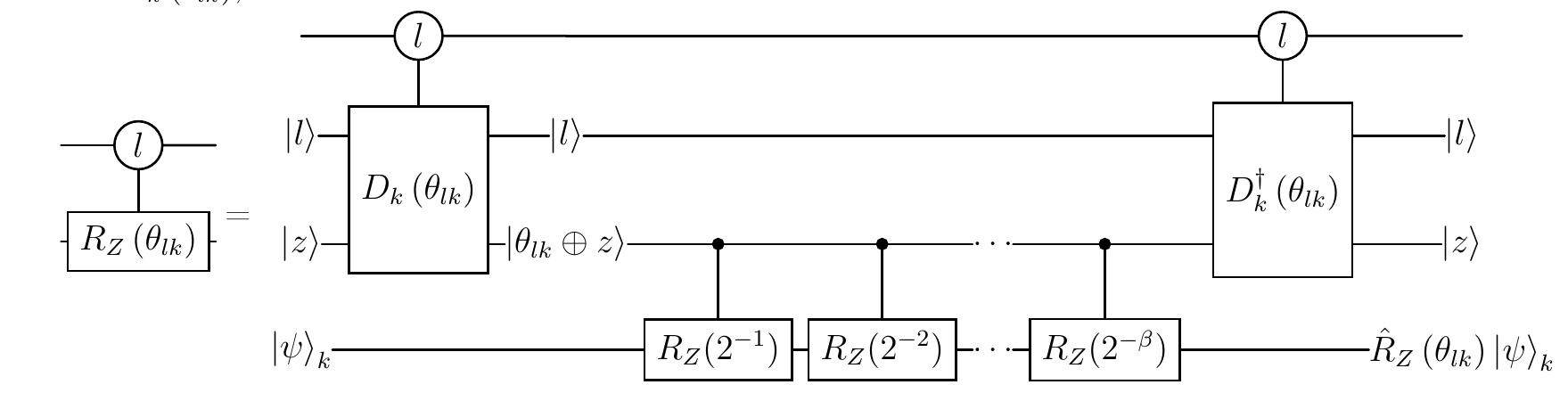}
	\caption{Quantum circuit representing multiplexed Z-rotation with data access oracles (QROMs) and fixed-angle Z-rotations, directly decomposable into Clifford+T gates with the method of Ref.~\cite{Gosset:2024}.}
	\label{fig:CR}
\end{figure}

\begin{figure}[!h] 
	\centering
	\includegraphics[width=0.7\linewidth]{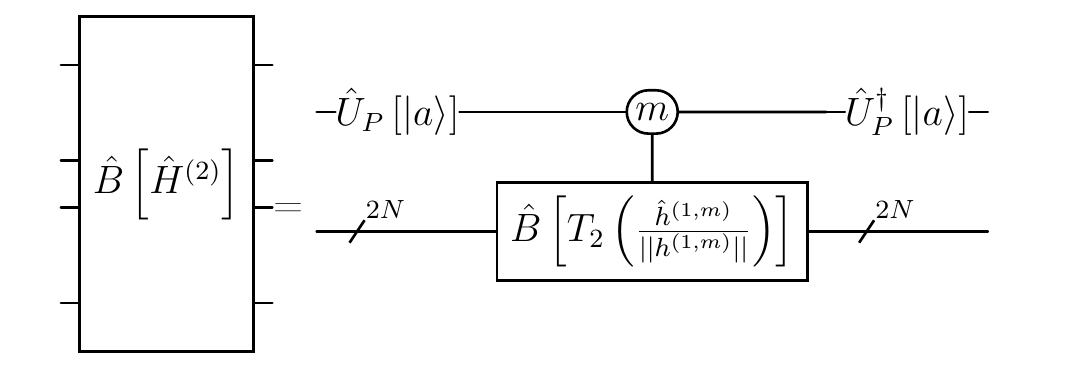}
	\caption{Quantum circuit representing block-encoding unitary for the two-body part of the Pauli-Breit Hamiltonian. Here $a_m = ||h^{(1,m)}||$ is the block-encoding scaling constant~\cite{Deka2025}.
	 }
	\label{fig:b2}
\end{figure}

\begin{figure}[!h] 
	\centering
	\includegraphics[width=0.7\linewidth]{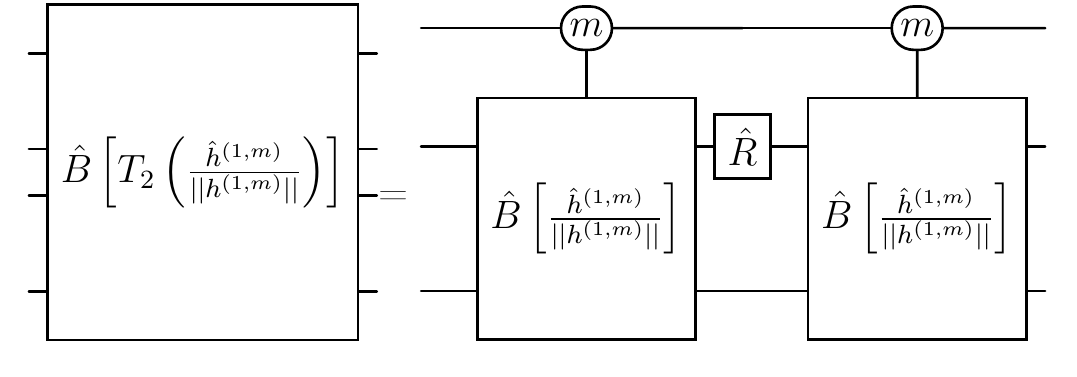}
	\caption{Qubitization circuit building the Chebyschev polynomial of one-body Hamiltonian - a unitary primitive for the LCU circuit shown in Fig.~\ref{fig:b2}. }
	\label{fig:b2a}
\end{figure}

Controlled $Z$ rotations with angles ${\theta_{l\mu k}}, {k=1,\ldots,2N-1}$ are implemented using data-access oracles $D_k\left(\theta{l\mu k}\right)$, as shown in Fig.~\ref{fig:CR}, together with single-qubit $R_Z(2^{-b})$ rotations by fixed angles $\pi/2^b$. The construction and associated cost of the data-access oracle $D_k\left(\theta_{l\mu k}\right)$ are given in Appendix~\ref{sec:appendixB}. The value register, initialized in the state $\ket{z}$ in the circuit of Fig.~\ref{fig:CR}, consists of $b$ qubits. The size of this precision register can be adjusted according to the desired numerical accuracy of the calculation.
Having block-encoded the one-body operator, block encoding of two-body part can be achieved with qubitization.
The block-encoding circuit for the two-body operator given in eq.~\ref{eq:two-body-chebyschev-qubitization2} is shown in Fig.~\ref{fig:b2}. The Chebyschev polynomial of the one-body Hamiltonian is implemented via qubitization as depicted in Fig.\ref{fig:b2a}.

\begin{figure}[!h] 
	\centering
	\includegraphics[width=\linewidth]{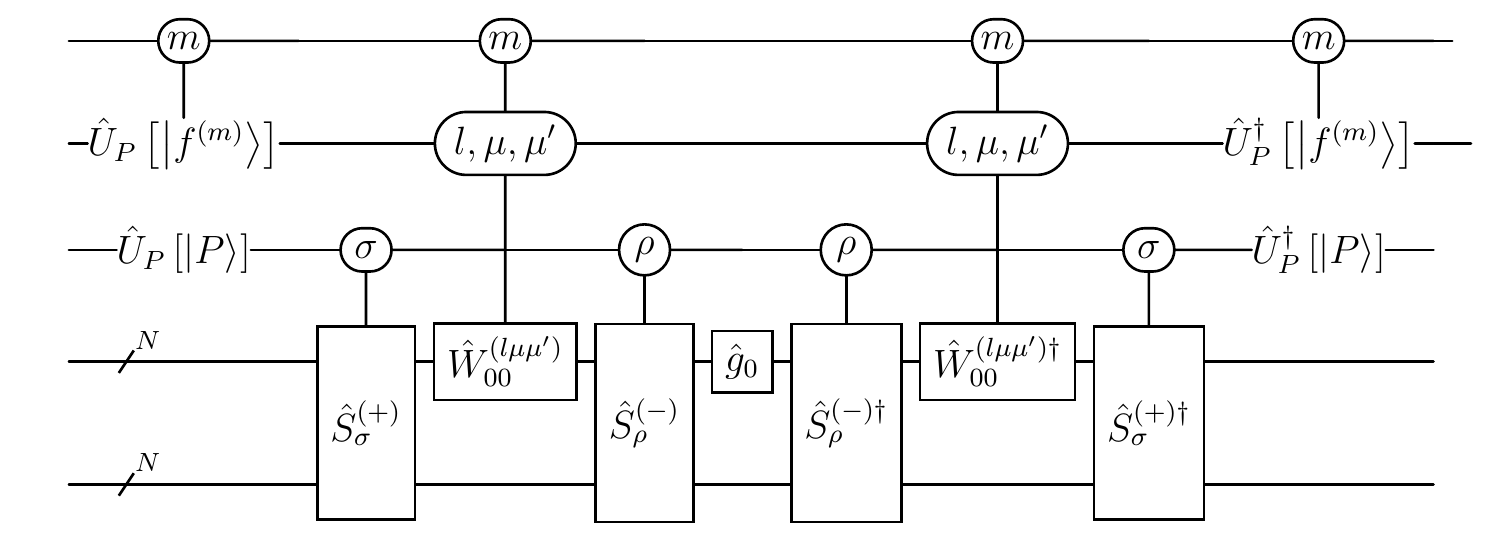}
	\caption{Quantum circuit representing block-encoding of the one-body operator within the block-encoding circuit of the two-body part in the PB Hamiltonian.  }
	\label{fig:b2-full}
\end{figure}

While we do not anticipate significant increase in the orbital basis size for solving the Schroedinger equation with the PB Hamiltonian, the spin-mixing operators enlarge the computational Hilbert space at least proportionally to the total spin of the system $\mathcal{O}(S)$. This factor can in prinicple become problematic, since non-relativistic electronic structure calculations already require substantial resources in terms of T-gates~\cite{Deka2025,loaiza2024,lee2021}. Minding this, our algorithm reduces the complexity of simulating the spin-mixed problem to that of the spin-independent problem, with little constant overhead.  

\section{Applications}

The development of new molecular compounds for photodynamic therapy (PDT) is strongly motivated by unmet needs in cancer treatment, where spatial selectivity, reduced systemic toxicity, and compatibility with combination therapies are known challenges~\cite{Agostinis2011,Wang2024,Chen2025}. PDT offers a unique therapeutic modality in which cytotoxic activity is confined to illuminated tumor regions, making it particularly attractive for localized and minimally invasive cancer treatments. However, the clinical performance of many existing photosensitizers is limited by insufficient triplet yields, poor photostability, inadequate tissue penetration due to suboptimal absorption wavelengths, or unfavorable pharmacokinetics~\cite{Dougherty1998,Wang2024}. These limitations directly trace back to molecular excited-state properties, especially the efficiency of intersystem crossing and subsequent reactive oxygen species generation. As a result, rational, physics-based design of new photosensitizers guided by quantitative understanding of spin-orbit coupling, excited-state energetics, and triplet-state dynamics has emerged as a key strategy for improving PDT outcomes in oncology. This view represents the general computer-aided drug discovery (CADD) paradigm. Advances in electronic-structure theory and excited-state simulations now enable predictive screening of candidate compounds, offering the prospect of accelerating the discovery of photosensitizers optimized for cancer-specific therapeutic requirements while reducing reliance on empirical trial-and-error.

For example, the penetration depth of long-wavelength near infrared (NIR) light is significantly higher than that of short-wavelength UV and visible light, and thus NIR light in the second window (NIR-II) is acknowledged as the preferred phototherapeutic means for eliminating deep-seated tumors, given the higher maximum permissible exposure, reduced phototoxicity and low autofluorescence, among others~\cite{Yang2024}. For designing new PDT therapeutics, it is thus necessary to carry out accurate simulations at managable cost.

\begin{figure}[!h] 
	\centering
	\includegraphics[width=\linewidth]{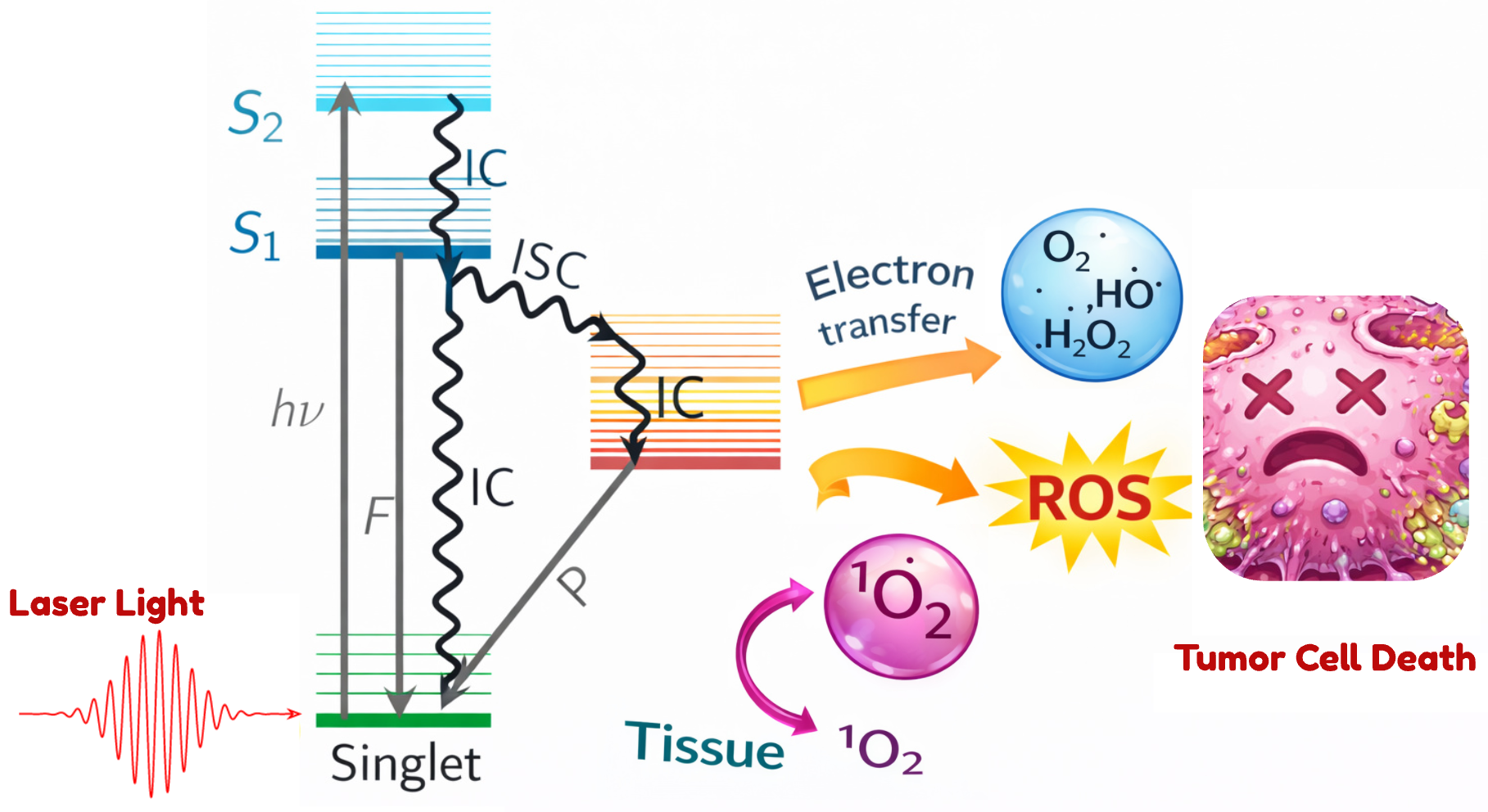}
	\caption{Schematic of intersystem crossing and reactive oxygen species generation for tumor cell destruction.  }
	\label{fig:isc}
\end{figure}

PDT relies on population transfer from optically excited singlet states to triplet manifolds, from which reactive oxygen species (ROS) are generated~\cite{Schweitzer2003}. The efficiency of this transfer is governed by intersystem crossing (ISC), a spin-forbidden process enabled by SOC~\cite{Marian2011,Marian2021}, shown schematically in Fig.~\ref{fig:isc}. At the microscopic level, SOC arises from relativistic corrections to the electronic Hamiltonian enclosed by the PB Hamiltonian.

Within a perturbative treatment, the ISC rate between an excited singlet state $\ket{S_i}$ and a triplet state $\ket{T_f}$ is given by Fermi’s golden rule,
\[
k_{\mathrm{ISC}}(S_i \rightarrow T_f)
=
\frac{2\pi}{\hbar}
\left|
\left\langle S_i \left| \hat{H}^{\mathrm{(SO)}} \right| T_f \right\rangle
\right|^2
\rho(E_{S_i}-E_{T_f}),
\]
where $\hat{H}^{\mathrm{(SO)}}$ denotes the spin-orbit coupling operator, which can be taken either as $ \hat{H}^{\mathrm{SO,1}}$ defined by eq.~\ref{eq:ham-SO1}, a sum of one-body and two-body spin-orbit coupling hamiltonians $ \hat{H}^{\mathrm{SO,1}}+\hat{H}^{\mathrm{SO,2}}$, or an effective spin-orbit interaction Hamiltonian~\cite{Epifanovsky2015},  and $\rho(E)$ is the vibronic density of states at the singlet-triplet energy gap. In the PB formalism, $\hat{H}^{\mathrm{(SO)}}$ emerges naturally from relativistic one-body spin-orbit terms as well as two-body spin-other-orbit and spin-spin contributions, enabling a first-principles description of SOC effects without empirical parametrization.

This explicit Hamiltonian-level connection is particularly relevant for photosensitizer design, where ISC efficiency is controlled by both the magnitude of SOC matrix elements and the energetic proximity of excited states~\cite{Penfold2018}. According to El-Sayed’s rules, ISC is enhanced when the singlet and triplet states differ in orbital character, a qualitative principle that is quantitatively captured by the matrix element
\[
\left\langle S_i \left| \hat{H}^{\mathrm{(SO)}} \right| T_f \right\rangle,
\]
which depends sensitively on orbital angular momentum, electron correlation, and relativistic effects encoded in the PB Hamiltonian.

From an order-of-magnitude perspective, SOC matrix elements scale approximately as $\xi \sim Z^4$ for atomic spin–orbit constants, where $Z$ is the nuclear charge. As a result, heavy-atom substitution (e.g., halogens, chalcogens, or transition-metal centers) can enhance SOC strengths from the $\sim 1$–$10\ \mathrm{cm^{-1}}$ range typical of light-atom organic chromophores to $\sim 10^2$–$10^3\ \mathrm{cm^{-1}}$. Inserting these values into the golden-rule expression yields ISC rates spanning several orders of magnitude, from $k_{\mathrm{ISC}} \sim 10^6\ \mathrm{s^{-1}}$ in weak-SOC systems to $k_{\mathrm{ISC}} \gtrsim 10^{10}\ \mathrm{s^{-1}}$ in strongly relativistic environments, assuming comparable vibronic densities. Importantly, substantial variations in ISC rates can also be achieved in light-atom systems by engineering orbital character and state degeneracies, even in the absence of heavy atoms.

In PDT, the resulting triplet population determines the efficiency of both Type II mechanisms, involving energy transfer to molecular oxygen and singlet-oxygen generation, and Type I mechanisms, where electron or hydrogen transfer from the triplet state produces radical species. In both cases, the competition between ISC, radiative decay, and nonradiative relaxation makes $k_{\mathrm{ISC}}$ a key design parameter linking relativistic electronic structure calculation to macroscopic photochemical outcome.

\paragraph{Quantum-algorithm motivation.}

Accurate evaluation of SOC matrix elements and ISC rates poses a significant computational challenge for classical electronic-structure methods, particularly for systems with strong electron correlation, dense excited-state manifolds, or explicit two-body relativistic interactions. The PB Hamiltonian contains nontrivial spin-dependent operators whose exact treatment adds computational overhead over non-relativistic calculation, becoming more rapidly classically intractable. This motivates the development of quantum algorithms capable of simulating relativistic Hamiltonians directly in second quantization. Notably, the Pt-BODIPY complex, shown in Fig.\ref{fig:workflow} presents a multi-reference electronic structure (due to d-orbital involvement and ligand field effects) that pushes the limits of classical methods, making it an suitable target for our high-accuracy quantum algorithm. 

Quantum simulation frameworks based on block encoding and qubitization provide a natural route to treating the PB Hamiltonian, with systematically controllable error. By enabling access to eigenstates and transition matrix elements of $\hat{H}_{\mathrm{PB}}$, such algorithms offer the prospect of computing SOC matrix elements $\left\langle S_i \left| \hat{H}_{\mathrm{SOC}} \right| T_f \right\rangle$ and, by extension, ISC rates from first principles, without relying on perturbative or mean-field approximations. In this sense, quantum algorithms do not merely accelerate existing workflows, but open a pathway to quantitatively reliable simulations of spin-forbidden photophysics in regimes that are currently inaccessible to classical computation.

\subsection{Proposal for a hybrid quantum-classical method for the design and accurate virtual screening of photodynamic therapy molecular candidates}

\subsubsection{Outline of the procedure}
Building on the discussion of PDT applications in the previous section, and assuming fault-tolerant quantum computing capabilities of future hardware, we propose a workflow that combines the quantum algorithm introduced in this work with existing classical computational techniques. The purpose is to deliver a tool for identifying efficient photosensitizer compounds for photodynamic therapy, based on the simulation of intersystem crossing rates.

\begin{figure}[!h] 
	\centering
	\includegraphics[width=0.9\linewidth]{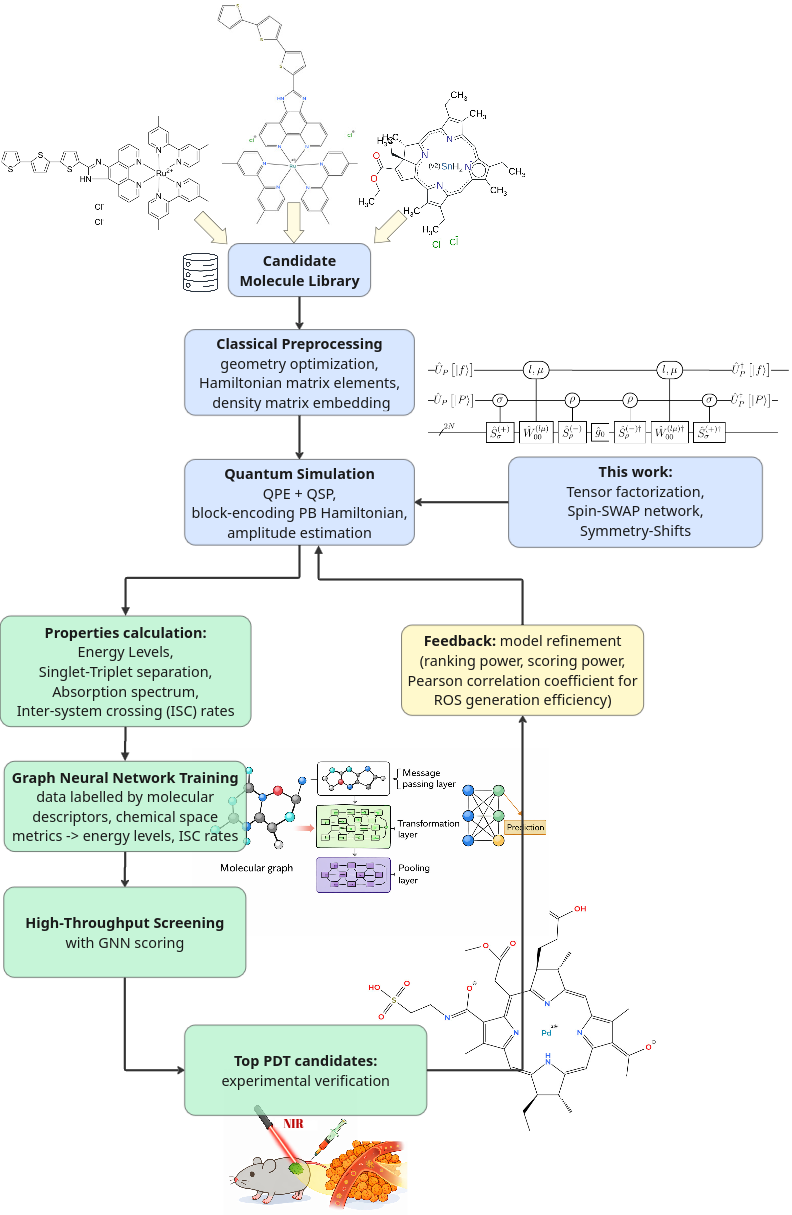}
	\caption{ High-level workflow of the hybrid quantum–classical pipeline for photodynamic therapy drug design. A fault-tolerant quantum algorithm computes key photophysical properties (singlet/triplet energies, spin–orbit couplings and intersystem crossing rates) for a selection of candidate molecules. These quantum-derived results then train a graph neural network model, which can rapidly predict properties for large libraries of new molecules. The GNN-guided screening identifies top candidates with desired absorption and ROS-generating characteristics. Subsequent experimental feedback helps establishing key metrics for quantum model refinement.}
	\label{fig:workflow}
\end{figure}

The process outlined in Fig.\ref{fig:workflow} begins by selecting candidate molecules such as Aza-BODIPYs, phthalocyanines, NIR-II chromophores, and transition metal complexes~\cite{Xu2022}, including heavy-atom substitutions that enhance SOC~\cite{Tram2009,Yang2024}. Each molecule undergoes classical geometry optimization (e.g. via DFT) and relativistic Hamiltonian construction, including all one- and two-electron SOC terms, to form the PB Hamiltonian $H_{\text{PB}}$ in mean-field spin-orbital basis. The Hamiltonian is mapped to qubits and block-encoded using our algorithm presented in sec.~\ref{sec:encoding}.

The quantum phase estimation algorithm with QSP is then used to compute the energy levels of $H_{\text{PB}}$, accurately resolving singlet and triplet levels (e.g. $E(S_0)$, $E(S_1)$, $E(T_1)$). To estimate SOC-driven ISC rates, off-diagonal matrix elements like $\langle S_1 | H_{\text{SO}} | T_1 \rangle$ are computed, either via Hadamard test or amplitude estimation~\cite{Rall2023}. ISC rates are approximated using Fermi’s Golden Rule: $k_{\text{ISC}} \propto |\langle S_1 | H_{\text{SO}} |T_1 \rangle|^2 / \Delta E^2$.

Quantum computing outputs for each molecule include: singlet/triplet excitation energies, SOC matrix elements, absorption wavelengths $\lambda$ and estimated ROS generation efficiency (via $k_{\text{ISC}}$ and $E(T_1)$). These high-fidelity quantum data points serve as training targets for a graph neural network (GNN) model. The model assumes certain radiation conditions.

Molecular graphs (atoms as nodes, bonds as edges) are used as input to a message-passing GNN with multitask heads to predict properties like $E(S_1)$, $E(T_1)$, SOC, and $\lambda_{\max}$. Heavy-atom features and 3D geometric features are included to capture relativistic effects and molecular fingerprint features. Molecular geometries and compositions are encoded in edges and nodes of the molecular graph. The GNN is trained on a few hundred to few thousand quantum-evaluated molecules, with transfer learning from classical or experimental databases where needed for further model refinement. The network can be expanded and adapted to accomodate other molecular metadata, including QSAR parameters.

Once trained, the GNN is deployed for high-throughput screening of a much larger virtual library, predicting SOC-driven properties and filtering compounds that maximize PDT efficacy (e.g. long-wavelength absorption, strong SOC, high $k_{\text{ISC}}$). This enables the design of novel PS candidates otherwise inaccessible through classical computation alone. Optionally, the model is iteratively refined via active learning, targeting molecules in uncertain regions of chemical space for additional quantum simulations.
Databases of compounds can be created through similarity generation algorithms using known classes of PDT agents collected in the literature, see e.g.~\cite{Wahnou2023}.

\subsubsection{Application to representative photosensitizers} 
\begin{figure}[!h] 
	\centering
	\includegraphics[width=0.9\linewidth]{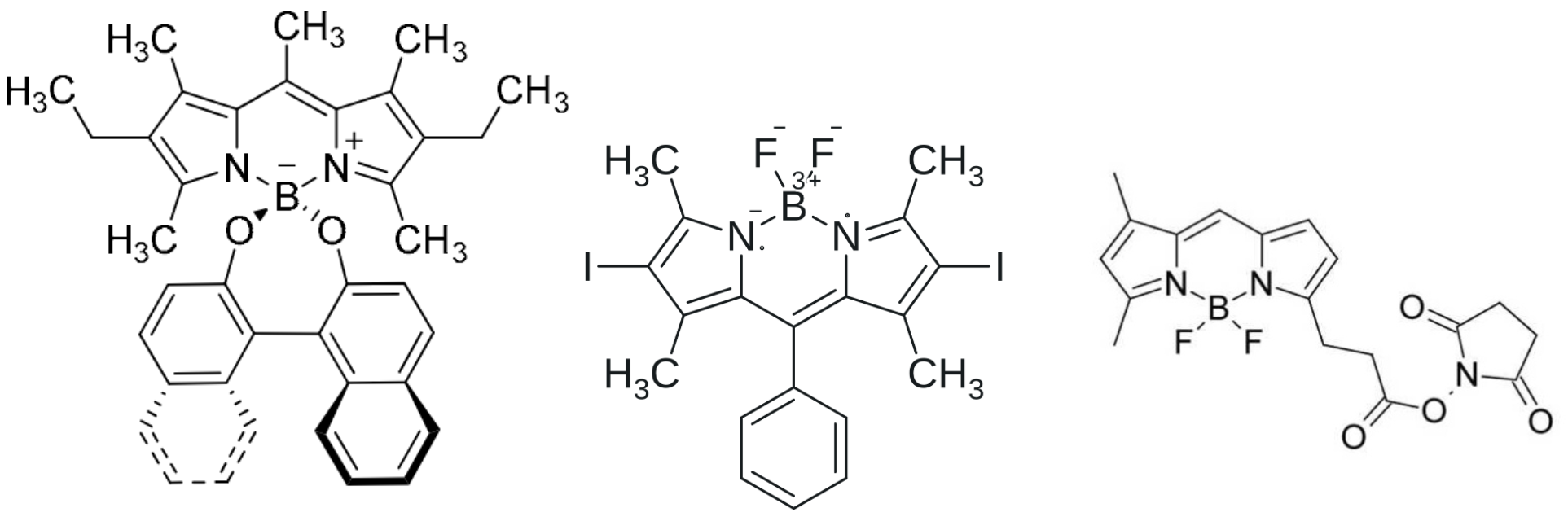}
	\caption{ Structures of BODIPY representatives. }
	\label{fig:example-PDTs}
\end{figure}

The quantum computation can be carried out on a diverse set of PDT compound classes, including Aza-BODIPY dyes, phthalocyanine macrocycles, and transition metal complexes, chosen for their relevance and challenging electronic structure. Example structures are shown in Fig.~\ref{fig:example-PDTs}. These systems span a range of heavy-atom substitution and $\pi$-extension approaches to the design for improving PDT performance. 
There are a number of clinically approved drugs for photodynamic therapy, including Porfimer
sodium, Protoporphyrin IX (Levulan), Temoporfin (Foscan), Verteporfin (Visudyne), Talaporfin
(Laserphyrin) and other~\cite{Kamkaew2013}. Padeliporfin, shown in Fig.~\ref{fig:workflow} is used to treat men with prostate cancer. Specific calculations of the PB matrix elements is beyond the scope of this work. Currently such a capability is not offered by any of the popular quantum chemistry packages. 

\subsubsection{Photophysical properties calculations} The quantum computations directly return key photophysical observables that determine PDT efficacy. In particular, two primary quantities are extracted for each candidate: (1) Absorption characteristics in the therapeutic window, and (2) Intersystem crossing efficiency. The photosensitizer’s light-harvesting ability is quantified by its cumulative absorption within the near-infrared tissue transparency window (~700–850 nm). 
The intersystem crossing rate is obtained by simulating the nonradiative spin conversion from the lowest excited singlet to triplet manifold. It can be estimated via a short-time time-evolution algorithm that propagates an initial excited singlet state under the spin–orbit-coupled Hamiltonian and measures the population transfer into a triplet state. The initial slope of singlet-to-triplet population versus time serves as a proxy for the ISC rate. It effectively captures the magnitude of spin-orbit coupling matrix elements between the states. A faster rise indicates more efficient ISC, which is desirable because only triplet excitons can sensitize molecular oxygen. Alternatively one can use amplitude estimation to ger directly the magnitude of the SOC matrix elements between excited singlet and triplet state.

\subsubsection{Graph Neural Network Model} To enable low-cost exploration of chemical space, we propose to integrate the above quantum computations with a machine learning model that generalizes the results to new compounds. The idea is to utilize the quantum-generated dataset (energies, oscillator strengths, ISC rates for representative molecules) as a training set for a predictive model, thereby creating a fast surrogate for property evaluation. In particular, one can train a graph neural network that takes a molecule’s 2D graph (atoms and bonds) as input and predicts its key photophysical parameters. Graph-based deep learning approaches have already shown success in modeling photosensitizer properties such as singlet oxygen quantum yields and absorption maxima, making them a natural choice for our surrogate~\cite{Wang2025}. 

GNN can be optimized to reproduce the quantum-computed observables for the training compounds, learning the subtle structure-property relationships. Once validated in experiment, this trained GNN model can rapidly screen thousands of candidate molecules, including those not explicitly simulated on quantum hardware, and estimate their absorption and ISC metrics with near-quantum accuracy. In this way, the workflow becomes a classical-quantum hybrid: we perform expensive but accurate quantum simulations on a strategically chosen subset of molecules, then the GNN interpolates those results across a broad chemical library at negligible computational cost. This improves scalability, allowing us to search chemical space far more efficiently than would be possible with direct quantum chemistry alone. Still, high-ranking new candidates proposed by the GNN can subsequently be verified with targeted quantum calculations, creating a closed-loop refinement if needed.

\section{Summary}
\label{sec:summary}
In this work, we have developed a fault-tolerant quantum algorithm for first-principles simulation of the Pauli-Breit Hamiltonian. Our main contribution is a constructive procedure for block-encoding the full PB Hamiltonian, including both spin-independent and spin-dependent one- and two-body terms, within the qubitization and quantum phase estimation framework. 

Building on tensor-network ideas and the Majorana representation of fermionic operators, we introduced a doubly factorized form of the PB Hamiltonian that is particularly well suited for quantum circuit implementation. This representation allows the Hamiltonian to be expressed as a linear combination of unitary operators with controlled scaling constants, and a reduced-cost block-encoding. We combined the spin-unitary group technique of ref.~\cite{Gould1993} with Hamiltonian block-encoding and proposed a circuit that decouples spin and spatial degrees of freedom. 
We thereby proposed an explicit algorithm for computing relativistic electronic energy levels using QPE, with resource requirements governed primarily by the block-encoding cost and target precision.

A key technical advance of this work is the construction of new quantum circuits for block-encoding spin-dependent one- and two-body interactions. By exploiting the Pauli-matrix structure of spin operators and introducing Spin-SWAP networks that decouple spin control from orbital control, we reduce the overhead associated with spin mixing. Compared to a direct application of previously proposed non-relativistic block-encoding techniques to the PB Hamiltonian, our approach gives a reduction in logical-level resources of up to a factor of two for the ground state energy estimation, while preserving favorable asymptotic scaling.

Beyond ground state calculations, the proposed framework naturally extends to the computation of electronically excited states through the use of spectral filtering and band-pass techniques within the block-encoding formalism~\cite{Lin2020} or landscape scanning methods~\cite{GRM25}. Furthermore, the formalism applies to arbitrary total spin manifolds by employing irreducible spherical tensor representations of spin-dependent operators. Finally, our construction is compatible with ongoing advances in quantum data-access oracles for further Clifford+T cost reduction. 

We motivated our work by the paradigm, in which quantum computers replace, at least in part, approximate semi-empirical models and experiment. Classical techniques ought to be then substituted by a quantum-computational technique offering error-controlled and first-principles solution, that can be systematically improved. The PB Hamiltonian is one such example. 
Another motivation is more pragmatic. We focused on molecular systems where relativistic spin effects play a decisive functional role. In particular, we discussed applications to photodynamic therapy, where the efficiency of intersystem crossing between singlet and triplet manifolds governs the generation of reactive oxygen species and ultimately therapeutic efficacy. Accurate prediction of these processes requires explicit treatment of one- and two-electron spin-orbit interactions, which are naturally captured within the Pauli-Breit Hamiltonian and are challenging to model reliably with classical approaches. Beyond PDT, the proposed framework is relevant to a broader class of photochemical and catalytic systems, including transition-metal complexes, photocatalysts for artificial photosynthesis, and materials exhibiting strong spin-orbit-induced relaxation pathways. 

Taken together, these results establish a pathway toward fault-tolerant quantum simulations of relativistic molecular Hamiltonians, with direct relevance to photochemistry, spin-dependent excited-state dynamics, and hybrid quantum-classical workflows for molecular design. The relative contribution from the quantum-computing solver can be gauged by the available logical qubit and T-gate resources of the day. We showed how these resources can be reduced approximately two-times compared to existing algorithms.
Further work thus will include integration with classical methods and case studies, with the aim of rational molecular design in regimes that are inaccessible to current classical electronic-structure methods.

\section{Acknowledgments}
The author thanks Konrad Deka for helpful discussions.
This work was funded by the European Innovation Council accelerator grant COMFTQUA, no. 190183782. 

\section{Appendix A: using the unitary group approach to represent the Pauli-Breit Hamiltonian}
\label{sec:AppendixA}
We define the action of electron spin operators on the standard basis in the following way:

\begin{eqnarray}
		\hat{s}_z\ket{\sigma}=\sigma\ket{\sigma},\;\sigma=\pm\frac{1}{2}\\
		\hat{s}^2_z\ket{\sigma}=\sigma(\sigma+1)\ket{\sigma},\;\sigma=\pm\frac{1}{2}\\
		\hat{s}_+\ket{\sigma=+\frac{1}{2}}\equiv\hat{s}_+\alpha = 0\\
		\hat{s}_+\ket{\sigma=-\frac{1}{2}}\equiv\hat{s}_+\beta = \alpha\\
		\hat{s}_-\ket{\sigma=-\frac{1}{2}}\equiv\hat{s}_-\beta = 0\\
		\hat{s}_-\ket{\sigma=+\frac{1}{2}}\equiv\hat{s}_-\alpha = \beta\\
\end{eqnarray}
The spherical tensor spin operators are related to Cartesian spin operators as follows:
\begin{eqnarray}
	\hat{s}_x = \frac{1}{2}\left(\hat{s}_++\hat{s}_-\right) \\
		\hat{s}_y = \frac{1}{2i}\left(\hat{s}_+-\hat{s}_-\right) \\
			\hat{s}_z = \hat{s}_0 \\
\end{eqnarray}
giving the following matrix elements:
\begin{eqnarray}
	\bra{\sigma}\hat{s}_x\ket{\rho} = \frac{1}{2}\left(\delta_{\sigma \alpha}\delta_{\rho \beta}+\delta_{\sigma \beta}\delta_{\rho \alpha}\right) \\
	\bra{\sigma}\hat{s}_y\ket{\rho}  = \frac{1}{2i}\left(\delta_{\sigma \alpha}\delta_{\rho \beta}-\delta_{\sigma \beta}\delta_{\rho \alpha}\right) \\
	\bra{\sigma}\hat{s}_z\ket{\rho}  =\frac{1}{2} \delta_{\sigma\rho}\left(\delta_{\sigma \alpha}-\delta_{\sigma \beta}\right) \\
\label{app:eq:spin-matelem}
\end{eqnarray}

\subsection{PB Hamiltonian details}
The electronic Hamiltonian can be expressed by the $\mathcal{U}(N)$ generators 
\begin{equation}
	\hat{H}^{(el.)} = \sum_{p,q=1}^N\bra{p}	\hat{T}^{(e)}+\hat{V}^{(ne)}) \ket{q}\hat{E}_{pq}+\frac{1}{2}\sum_{p,q,r,s=1}^N\bra{pq}	\hat{V}^{(ee)} \ket{rs}\hat{E}_{pqrs}
	\label{eq:ham-el-2nd-quant}
\end{equation}
where
\begin{equation}
	\hat{E}_{pqrs} = \hat{E}_{pq}\hat{E}_{rs}-\delta_{rq}\hat{E}_{ps}=\sum_{\sigma,\rho}\hat{a}^{\dagger}_{p\sigma}\hat{a}_{q\sigma}\hat{a}^{\dagger}_{r\rho}\hat{a}_{s\rho}-\delta_{rq}\sum_{\sigma}\hat{a}^{\dagger}_{p\sigma}\hat{a}_{s\sigma}
\end{equation}
In eq.~\ref{eq:ham-el-2nd-quant}, $\hat{T}^{(e)}$ stands for the total electronic kinetic energy,  $\hat{V}^{(ne)}$ is the nuclear-electron attraction term and $\hat{V}^{(ee)}$ is the electron-electron interaction term, as defined in eq.~\ref{eq:ham-el}. 
The one-body Darwin term is given by:
\begin{equation}
	\hat{H}^{(D1)} = \sum_{p,q=1}^N\bra{p}	\hat{H}^{(D1)} \ket{q}\hat{E}_{pq}
	\label{eq:ham-D1-2nd-quant}
\end{equation}
The one-body spin-orbit terms are given by:
\begin{equation}
	\hat{H}^{(SO,1)} =\frac{1}{2}\sum_{\sigma,\rho} \mathbf{P}_{\sigma\rho} \cdot \sum_{p,q=1}^N  \bra{p}	\hat{\mathbf{H}}^{(Orb,1)} \ket{q}\hat{E}_{p\sigma q\rho}
	\label{eq:ham-SO1-2nd-quant}
\end{equation}
where $ \mathbf{P} $ is a vector of Pauli matrices $\mathbf{X},\mathbf{Y},\mathbf{Z}$ with elements $ \mathbf{P}^{(\mu)}_{\sigma\rho}$.
The triplet excitation operators are defined as
\begin{eqnarray}
	\mathbf{T}_{pq}^{(\mu)} = \sum_{\sigma,\rho} \hat{P}^{(\mu)}_{\sigma\rho}\hat{E}_{p\sigma q\rho}
	\label{eq:triple-excitation-op}
\end{eqnarray}
and represent the $SU(2)$-symmetry-adapted form of the $U(2N)$ Lie group generators $\hat{E}_{p\sigma q\rho}$. At the same time, the triplet excitation operators represent rank-1 tensor operators over the spin $SU(2)$ space, on account of the commutation relation with the spin operators: $\left[\hat{s}_{\mu},T_{pq}^{(\nu)}\right]_- = i\epsilon_{\mu\nu\tau}\hat{s}_{\tau}, \; p,q=1,2,...,N$. With these triplet excitation operators, the one-body spin-orbit Hamiltonian is written as:
\begin{equation}
	\hat{H}^{(SO,1)} =  \sum_{p,q=1}^N \mathbf{T}_{pq} \cdot  \bra{p}	\hat{\mathbf{H}}^{(Orb,1)} \ket{q}
	\label{eq:ham-SO1-2nd-quant-triplet}
\end{equation}
where 
\begin{equation}
	\hat{\mathbf{H}}^{(Orb,1)} =  \frac{1}{2}\sum_{\alpha=1}^{M}\frac{Z_{\alpha}}{||\mathbf{r_j}-\mathbf{R_{\alpha}}||^3}\left[\mathbf{l}_j-(\mathbf{R_{\alpha}}\times \mathbf{p_j})\right]
	\label{eq:ham-orb-1}
\end{equation}
is the vector operator representing the orbital angular momentum part of the one-body spin-orbit Hamiltonian.
The two-body spin-orbit coupling terms are given by:
\begin{equation}
	\hat{H}^{(SO,2)} =\frac{1}{4}\sum_{\sigma,\rho} \mathbf{P}_{\sigma\rho} \cdot \sum_{p,q,r,s=1}^N  \bra{pq}\hat{\mathbf{H}}^{(Orb,2)} \ket{rs}\left(\hat{E}_{p\sigma r\rho}\hat{E}_{qs}+\hat{E}_{pr}\hat{E}_{q\sigma s\rho}-2\delta_{qr}\hat{E}_{p\sigma s\rho}\right)
	\label{eq:ham-SO2-2nd-quant}
\end{equation}
and expressed with triplet excitation operators:
\begin{equation}
	\hat{H}^{(SO,2)} = \frac{1}{2} \sum_{p,q,r,s=1}^N \bra{pq}\hat{\mathbf{H}}^{(Orb,2)} \ket{rs} \cdot\left[ \mathbf{T}_{pq}\hat{E}_{qs}+\hat{E}_{pr}\mathbf{T}_{qs}-2\delta_{qr}\mathbf{T}_{ps} \right]
	\label{eq:ham-SO2-2nd-quant-triplet}
\end{equation}
where

\begin{eqnarray}
	\hat{\mathbf{H}}^{(Orb,2)} =\frac{1}{||\mathbf{r_j}-\mathbf{r_k}||^3}\cdot\left[(\mathbf{r}_j-\mathbf{r_k})\times (\mathbf{p_j}-\mathbf{p_k})\right]
\end{eqnarray}
is the vector operator representing orbital angular momentum coupling between different electrons. 
The spin-spin interaction term is given as:

\begin{align}
	\hat{H}^{(SS)} = \frac{1}{2} \sum_{p,q,r,s=1}^N \sum_{\sigma,\rho} \sum_{\nu,\tau} \mathbf{P}_{\sigma\rho} \cdot
	\bra{pq}\hat{\mathbf{h}}^{(SS)} \ket{rs}\cdot \mathbf{P}_{\nu\tau}\left(\hat{E}_{pr,\sigma\tau}\hat{E}_{qs,\rho\nu} - \delta_{qr}\delta_{\rho\tau}\hat{E}_{ps,\sigma\nu}\right)+\\+\frac{1}{8} \sum_{p,q,r,s=1}^N	\bra{pq}\hat{\mathbf{h}}^{(contact)} \ket{rs}\left(2\hat{E}_{qprs}+\hat{E}_{pqrs}\right)
	\label{eq:ham-SS-2nd-quant}
\end{align}
and expressed with triplet excitation operators:
\begin{equation}
	\hat{H}^{(SS)} = \frac{1}{2} \sum_{p,q,r,s=1}^N \mathbf{T}_{pr} \cdot
	\bra{pq}\hat{\mathbf{h}}^{(SS)} \ket{rs}\cdot \mathbf{T}_{qs}-\frac{1}{8} \sum_{p,q,r,s=1}^N	\bra{pq}\hat{\mathbf{h}}^{(contact)} \ket{rs}\left(2\hat{E}_{qpGosset:2024rs}+\hat{E}_{pqrs}\right)
	\label{eq:ham-SS-2nd-quant-triplet}
\end{equation}

\subsection{Appendix A.1. Derivation of the Majorana form for the one-body spin-orbit Hamiltonian.}
Recall eq.~\ref{eq:one-body-full-triplet}:
\begin{eqnarray}
	\hat{H}^{(SO,1)} = \sum_{p,q=1}^N\sum_{\mu = X,Y,Z} H^{(SO,1,\mu)}_{p,q}\hat{T}^{(\mu)}_{pq} \equiv \sum_{\mu = X,Y,Z}	\hat{H}^{(SO,1,\mu)}
	\label{A1:def}
\end{eqnarray}
in the Majorana representation, the respective triplet excitation components are given as
\begin{align}
\hat{T}^{(X)}_{pq} = \frac{1}{8}\left[\hat{\gamma}_{p\alpha 0}\hat{\gamma}_{q\beta 0}+\hat{\gamma}_{p\alpha 1}\hat{\gamma}_{q\beta 1}+\hat{\gamma}_{p\beta 0}\hat{\gamma}_{q\alpha 0}+\hat{\gamma}_{p\beta 1}\hat{\gamma}_{q\alpha 1}+i\left(\hat{\gamma}_{p\alpha 0}\hat{\gamma}_{q\beta 1}+\hat{\gamma}_{p\beta 0}\hat{\gamma}_{q\alpha 1}-\hat{\gamma}_{p\beta 1}\hat{\gamma}_{q\alpha 0}-\hat{\gamma}_{p\alpha 1}\hat{\gamma}_{q\beta 0}\right)\right] \\
\hat{T}^{(Y)}_{pq} = \frac{1}{8i}\left[\hat{\gamma}_{p\alpha 0}\hat{\gamma}_{q\beta 0}+\hat{\gamma}_{p\alpha 1}\hat{\gamma}_{q\beta 1}-\hat{\gamma}_{p\beta 0}\hat{\gamma}_{q\alpha 0}-\hat{\gamma}_{p\beta 1}\hat{\gamma}_{q\alpha 1}+i\left(\hat{\gamma}_{p\alpha 0}\hat{\gamma}_{q\beta 1}+\hat{\gamma}_{p\beta 1}\hat{\gamma}_{q\alpha 0}-\hat{\gamma}_{p\alpha 1}\hat{\gamma}_{q\beta 0}-\hat{\gamma}_{p\beta 0}\hat{\gamma}_{q\alpha 1}\right)\right] \\
\hat{T}^{(Z)}_{pq} = \frac{1}{8}\left[\hat{\gamma}_{p\alpha 0}\hat{\gamma}_{q\alpha 0}+\hat{\gamma}_{p\alpha 1}\hat{\gamma}_{q\alpha 1}-\hat{\gamma}_{p\beta 0}\hat{\gamma}_{q\beta 0}-\hat{\gamma}_{p\beta 1}\hat{\gamma}_{q\beta 1}+i\left(\hat{\gamma}_{p\alpha 0}\hat{\gamma}_{q\alpha 1}+\hat{\gamma}_{p\beta 1}\hat{\gamma}_{q\beta 0}-\hat{\gamma}_{p\beta 0}\hat{\gamma}_{q\beta 1}-\hat{\gamma}_{p\alpha 1}\hat{\gamma}_{q\alpha0}\right)\right] 
	\label{A1:triplet}
\end{align}
We then split the summation over orbital indices into the upper triangle, the lower triangle and the diagonal part as follows:
\begin{equation}
	\hat{H}^{(SO,1)} =\sum_{\mu = X,Y,Z} \left[\sum_{p>q} H^{(SO,1,\mu)}_{p,q}\hat{T}^{(\mu)}_{pq}+\sum_{p<q} H^{(SO,1,\mu)}_{p,q}\hat{T}^{(\mu)}_{pq}+\sum_{p} H^{(SO,1,\mu)}_{p,p}\hat{T}^{(\mu)}_{pp}\right]
	\label{A1:split}
\end{equation}
and use the following facts:
\begin{enumerate}
	\item  the $X$-component of the orbital angular momentum operator generates a symmetric representation, i.e. $H^{(SO,1,X)}_{p,q} = H^{(SO,1,X)}_{q,p}$ and the $X$-component of the triplet excitation operator is symmetric with respect to index exchange.
	\item  the $Y$-component of the orbital angular momentum operator generates a hermitian representation, i.e. $H^{(SO,1,Y)}_{p,q} = H^{(SO,1,Y)}*_{q,p}$ and the $Y$-component of the triplet excitation operator is anti-symmetric with respect to index exchange. In other words $H^{(SO,1,Y)}_{p,q} = i\tilde{H}^{(SO,1,Y)}_{p,q}$, where $\tilde{H}^{(SO,1,Y)}_{p,q}$ is a skew-symmetric (antisymmetric) matrix.
	\item  the $Z$-component of the orbital angular momentum operator generates a diagonal representation, i.e. $H^{(SO,1,Z)}_{p,q} = \delta_{pq}H^{(SO,1,Z)}_{p,q}$ and the $Z$-component of the triplet excitation operator is anit-symmetric with respect to index exchange.
\end{enumerate} 
such that the $X$-component of the spin-orbit one-body Hamiltonian can be written as:
\begin{eqnarray}
	\hat{H}^{(SO,1,X)} = \frac{i}{4}\sum_{p,q} H^{(SO,1,X)}_{p,q}\left(\hat{\gamma}_{p\beta 0}\hat{\gamma}_{q\alpha 1}+\hat{\gamma}_{p\alpha 0}\hat{\gamma}_{q\beta 1}\right)
	\label{A1:split-2}
\end{eqnarray}
where we used the Majorana commutation relations. The real part of the $X$-compoment triplet excitation operator given in eq.~\ref{A1:triplet} vanishes. We can express eq.~\ref{A1:split-2} with the use of $SU(2)$ generator $P^{(X)}_{\sigma\rho}$ for the spin index, to arrive at:
\begin{equation}
	\hat{H}^{(SO,1,X)} = \frac{i}{4}\sum_{p,q=1}^N H^{(SO,1,X)}_{p,q}\sum_{\sigma,\rho}\hat{\gamma}_{p\sigma 0}P^{(X)}_{\sigma\rho}\hat{\gamma}_{q\rho 1}
\end{equation}

For the $Y$-component, we first note that the matrix elements form a hermitian matrix:
\begin{eqnarray}
H^{(SO,1,Y)}_{p,q} = \int_{\mathbb{R}^3}\phi^*_{p}(r)\xi(r)\hat{L}_Y\phi_{q}(r)d^3\mathbf{r} = H^{(SO,1,Y)}*_{q,p}
\end{eqnarray} 
such that the decomposition given in eq.~\ref{A1:split-2} reads
\begin{eqnarray}
	\hat{H}^{(SO,1,Y)} =\sum_{p>q} H^{(SO,1,Y)}_{p,q}\hat{T}^{(Y)}_{pq}+H^{(SO,1,Y)}*_{p,q}\hat{T}^{(Y)}_{qp}+\sum_{p} H^{(SO,1,Y)}_{p,p}\hat{T}^{(Y)}_{pp}
	\label{A1:split-Y}
\end{eqnarray}
because the representation of the $\hat{L}_Y$ operator are purely imaginary we get:
\begin{eqnarray}
	\hat{H}^{(SO,1,Y)} =\sum_{p>q} H^{(SO,1,Y)}_{p,q}\left(\hat{T}^{(Y)}_{pq}-\hat{T}^{(Y)}_{qp}\right)+\sum_{p} H^{(SO,1,Y)}_{p,p}\hat{T}^{(Y)}_{pp}
	\label{A1:split-Y-2}
\end{eqnarray}
which leads to a complex representation:
\begin{equation}
	\hat{H}^{(SO,1,Y)} = \frac{i}{4}\sum_{p,q=1}^N\sum_{\sigma\rho} H^{(SO,1,Y)}_{p,q}\hat{P}^{(Y)}_{\sigma\rho}\left[ (1-\delta_{pq})\theta(q-p)\hat{\gamma}_{p\sigma 0}\hat{\gamma}_{q\rho 1}+\frac{i}{2}\delta_{pq}(\hat{\gamma}_{p\sigma 0}\hat{\gamma}_{p\rho 0}+\hat{\gamma}_{p\sigma 1}\hat{\gamma}_{p\rho 1})\right]
	\label{A1:y-comp}
\end{equation}

although the imaginary diagonal part in eq.~\ref{A1:y-comp} can be small or vanishing by symmetry. Note that Pauli-Y is a matrix with purely imaginary components. The coefficients in eq.~\ref{A1:y-comp} can be included into matrix elements $H^{(SO,1,Y)}_{p,q}$ to give:
\begin{equation}
	\hat{H}^{(SO,1,Y)} = \frac{i}{4}\sum_{p\neq q=1}^N\sum_{\sigma\rho} \tilde{H}^{(SO,1,Y)}_{p,q}\hat{\gamma}_{p\sigma 0}\hat{P}^{(Y)}_{\sigma\rho}\hat{\gamma}_{q\rho 1}+\frac{1}{4}\sum_{p=1}^N H^{(SO,1,Y)}_{p,p}\sum_{\sigma\rho}\frac{1}{2}(\hat{\gamma}_{p\sigma 0}\hat{P}^{(Y)}_{\sigma\rho}\hat{\gamma}_{p\rho 0}+\hat{\gamma}_{p\sigma 1}\hat{P}^{(Y)}_{\sigma\rho}\hat{\gamma}_{p\rho 1})
	\label{A1:y-comp-2}
\end{equation}
where $ \tilde{H}^{(SO,1,Y)}_{p,q} =  H^{(SO,1,Y)}_{p,q}\theta(q-p)$.
The $Z$-component $	\hat{H}^{(SO,1,Z)}$ can be calculated with analogous reasoning as the $X$-component.
Putting all components together we get the one-body spin-orbit Hamiltonian written as:
\begin{equation}
	\hat{H}^{(SO,1)} = \frac{i}{4}\left[\sum_{p,q=1}^N \sum_{\sigma\rho} \sum_{\mu = X,Y,Z} H^{(SO,1,\mu)}_{p,q}
	\hat{\gamma}_{p\sigma 0}\hat{P}^{(\mu)}_{\sigma\rho}\hat{\gamma}_{q\rho 1}-\sum_{p=1}^N \sum_{\sigma\rho} \sum_{s,t} H^{(SO,1,Y)}_{p,p}\left(\hat{\gamma}_{p\sigma s}\hat{P}^{(X)}_{\sigma\rho}\hat{\gamma}_{p\rho t}\left(\mathbf{1}_{st}+\hat{X}_{st}\right)\right)\right]
	\label{A1:all-comp}
\end{equation}

\subsection{Derivations for the factorized Hamiltonian}
Adopting the spin-dependent decomposition scheme of eq.~\ref{eq:h1-decomposition-spin}, the one-body Hamiltonian can be written as:
\begin{eqnarray}
	\hat{H}^{(1)} = \frac{i}{4}\sum_{l=1}^{l_{max}}\tilde{f}^{(1)}_l\left(\sum_{p=1}^N\sum_{\sigma}v_{l p\sigma}\hat{\gamma}_{p\sigma 0}\right)\cdot\left(\sum_{q=1}^N\sum_{\rho} v^{T}_{l q\rho}\hat{\gamma}_{q\rho 1}\right) \equiv  \frac{i}{4}\sum_{l=1}^{l_{max}}\tilde{f}^{(1)}_l\vec{\gamma}_{l0}\cdot \vec{\gamma}_{l1} 
	\label{eq:one-body-ham-majorana-decomposed}
\end{eqnarray}
where we defined a new vector spanned by the basis of Majorana operators:
\begin{eqnarray}
	\vec{\gamma}_{lx} = \sum_{p,\sigma}  v_{l p\sigma} \gamma_{p\sigma x}
	\label{vector-in-majora-operators-2N}
\end{eqnarray}
Our aim is to construct  a vector of Majorana operators that is both unitary and self-inverse, hereafter referred to as the Majorana vector. In this case, the corresponding block encoding is straightforward. From the perspective of quantum-computer implementation, the operators embedded in the unitary circuits executed on the quantum hardware take the following general form:
\begin{eqnarray}
	\hat{\Gamma}_{l}= \vec{\hat{\gamma}}_{l}\cdot\vec{\hat{\gamma}}_{l}
	\label{eq:gamma}
\end{eqnarray}
where
\begin{eqnarray}
	\vec{\gamma}_{l} = \sum_{p,\sigma,s}  v_{l p\sigma,s} \gamma_{p\sigma s}
	\label{vector-in-majora-operators-4N}
\end{eqnarray}
represents the Majorana vector of size $4N$. For the Pauli-Breit class of Hamiltonians it is convenient to define Majorana vectors of size $2N$, as shown in eq.~\ref{vector-in-majora-operators-2N}, in which case we have
\begin{eqnarray}
	\hat{\Gamma}_{l}= \ket{0}\bra{1}\otimes \vec{\hat{\gamma}}_{l0}\cdot\vec{\hat{\gamma}}_{l1}
	\label{eq:gamma-2N-spin}
\end{eqnarray}
Alternatively, one may want to represent the Majorana vectors for each spin value separately, in which case:
\begin{eqnarray}
	\hat{\Gamma}_{l}= \sum_{\sigma,\rho} \ket{\sigma}\bra{\rho} \otimes \ket{0}\bra{1}\otimes \vec{\hat{\gamma}}_{l\sigma 0}\cdot\vec{\hat{\gamma}}_{l\rho 1}
	\label{eq:gamma-N-spin}
\end{eqnarray}
The advantage of using representation given in eq.~\ref{eq:gamma-N-spin} over that of eq.~\ref{eq:gamma-2N-spin} becomes apparent throughout the block-encoding procedure of the electronic Hamiltonian, discussed sec.~\ref{sec:encoding}.
For spin-independent Hamiltonians eq.~\ref{eq:gamma} reduces to:
\begin{eqnarray}
	\hat{\Gamma}_{l}=\sum_{\sigma} \ket{\sigma}\bra{\sigma}\otimes \ket{0}\bra{1}\otimes \vec{\hat{\gamma}}_{l\sigma 0}\vec{\hat{\gamma}}_{l\sigma 1}
\end{eqnarray}
where the vectors of Majorana operators are of length $N$, i.e.: $	 \vec{\gamma}_{l\sigma x} = \sum_{p}  v_{l p} \gamma_{p\sigma x}$.

With spin-dependence explicitly moved to Pauli matrix representation the two-body Hamiltonian can be expressed as:
\begin{equation}
	\hat{H}^{(2)} = \sum_{m=1}^M \left(\sum_{l=1}^{L(m)} \sum_{\mu,\mu'=0,X,Y,Z} f^{(m,l,\mu,\mu')}\sum_{\sigma,\rho}\hat{P}^{(\mu)}_{\sigma\rho}\left(\sum_{p}v_{p,0}^{(m,l,\mu,\mu')}\gamma_{p\sigma 0}\right)\left(\left(\sum_{q}v_{q,1}^{(m,l,\mu,\mu')}\right)^T\gamma_{q\rho 1}\right)\right)^2 +\hat{\epsilon}\left(M,\mathbf{L}\right)
	\label{eq:two-body-DF-spin-out}
\end{equation}

Following reasoning similar to that presented by von Burg et al.~\cite{Burg2021}, we further require that the vector of Majorana operators be represented as a product of unitary operators, as given below:
\begin{eqnarray}
	\vec{\gamma}_{l,m} = \mathbf{U}^{(m,l)}\left(\mathbf{\theta}\right)\gamma_0\left(\mathbf{U}^{(m,l)}\left(\mathbf{\theta}\right)\right)^{\dagger}
	\label{unitary-to-unitary}
\end{eqnarray}
where $\gamma_0$ is seed operator, usually chosen as $\gamma_{000}$. The unitary transforming the Majorana vector into a product form can be written as:
\begin{eqnarray}
	\mathbf{U}^{(m,l)}\left(\mathbf{\theta}\right)= \prod_{k=1}^{4N-1}\hat{u}^{(m,l)}_{k}
	\label{eq:U-ansatz-aarhus-gauge}
\end{eqnarray}
where
\begin{eqnarray}
	\hat{u}^{(m,l)}_{k} = \exp\left(\theta_{mlk}\hat{\gamma}_{p(k)\sigma(k) x}\hat{\gamma}_{p(k+1)\sigma(k+1) x}\right)
	\label{eq:U-ansatz-aarhus-atom-exp}
\end{eqnarray}
where $p(k),\sigma(k)$ are mapping functions that distribute the action of the Majorana operators over the qubits. The decomposition shown in eq.~\ref{eq:U-ansatz-aarhus-gauge} is possible because the matrices formed from the coefficients $ v^{(m,l)}_{p'\sigma' x'}$ are rank-1, that is, they possess a single non-zero eigenvalue.

\section{Appendix B: Data-access oracles}
\label{sec:appendixB}
\subsection{Data-access oracle}
\begin{figure}
	\begin{quantikz}[column sep=0.2cm]
		\ket{j}	^{\otimes n_D}  &  \measure{j} & \\
		\ket{0} 				& \ctrl{1}\wire[u]{q}& \\
		\ket{0}^{\otimes \beta}	& \gate{U_s} &
	\end{quantikz}
	$\equiv$ \begin{quantikz}[column sep=0.08cm]
		\ket{j}	^{\otimes n_D}  &  \measure{j=0} 			& 					& \measure{j=0} 	&\measure{j=1}		& 				&\measure{j=1} 	& 						& \ldots & &\measure{j=N-1} & \ket{j}  \\
		\ket{0} 				& \targ{} \wire[u]{q} 	& \ctrl{1}			& \targ{} \wire[u]{q} & \targ{} \wire[u]{q}	& \ctrl{1}		&  \targ{} \wire[u]{q} &&\ldots  	& \ctrl{1}&  \targ{} \wire[u]{q}   &\ket{0}\\
		\ket{0}^{\otimes \beta}	& 						& \gate{X^{D(0)}} 	& 					&	& \gate{X^{D(1)}} 	& && \ldots & \gate{X^{D(N-1)}} & &\ket{D(j)}
	\end{quantikz}
	\caption{Schematic representation of the selection operator defined in eq.~\ref{DA:select}. Here $n_D = \lceil \log N\rceil$}
	\label{fig:DA-select}
\end{figure}
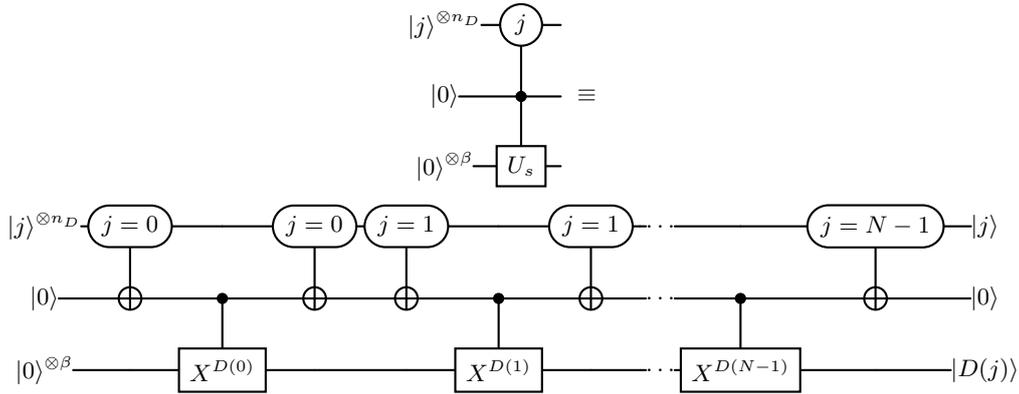

The block-encoding procedure requires an implementation of two operators: state preparation $V_P$ and unitary selection operator $V_S$. The former synthesizes a normalized quantum state with amplitudes encoding coefficients of the block-encoded Hamiltonian. The latter selects respective unitary components in the Hamiltonian's linear combination of unitaries representation. For construction of both $V_P$ and $V_S$ one must load information about the coefficients' identifiers in the linear combination and their values into quantum registers and state amplitudes, respectively. Data-access oracles load classical information into quantum registers, which can be written as
\begin{equation}
	\hat{D}\ket{j}\ket{0}=\ket{j}\ket{D(j)}
	\label{DA:def}
\end{equation}
where $\ket{j}\in \mathbb{C}^N$ is the argument state. A data-access oracle, as defined in eq.~\ref{DA:def}, is commonly constructed via a unitary operator that projects onto appropriately encoded size $N$ space of classical data arguments and applies an information-encoding unitary:
\begin{equation}
	\hat{U}_S = \sum_{j=0}^{N-1} \ket{j}\bra{j}\otimes \hat{U}_j \\
	\label{DA:select}
\end{equation}
where $ \hat{U}_j = \otimes_{k=0}^{\beta-1}\hat{X}^{d_{jk}}$ with $d_{jk}\in \lbrace 0,1\rbrace$ and $\hat{X}$ is the Pauli X gate. Registers $\ket{j}$ encode the arguments of the classical input data and $\beta$ gives the number of bits representing each classical entry. The operator defined in eq.~\ref{DA:select} is schematically depicted in Fig.~\ref{fig:DA-select}.

\begin{figure}
	\begin{quantikz}[column sep=0.05cm]
		\ket{a}  		&  \ctrl{4}   	& \\
		\ket{\phi_0} 	& \targ{}		& \\
		\ket{\phi_1}	& \targ{}		& \\
		\vdots			& \vdots		& \\
		\ket{\phi_{\lambda-1}}	& \targ{}		& 
	\end{quantikz}=\begin{quantikz}[column sep=0.5cm]
		\ket{a}  \slice{1}				&   \slice{2}  			&  	 \slice{3}  	&  \slice{4}  	 						& 	 \slice{5} 			&\ldots 	 \slice{$\log\lambda$}  			&\ctrl{1} \slice{}  		&\slice{} \ldots 		&		\slice{}	&  \slice{}			 	& \slice{2$\log\lambda$+1} & \slice{}  & \\
		\ket{\phi_0} 			& \ctrl{1}		&  \ctrl{2} 	& \ctrl{4}				& \ctrl{8} 	& \ldots	&\targ{} 	& \ldots 	&\ctrl{8}	& \ctrl{4}	& \ctrl{2}	&	\ctrl{1} 	& \\
		\ket{\phi_1}			& \targ{}		&  				& 						& 			&			&			& 			&			&			& &\targ{}		& \\
		\ket{\phi_2} 			& \ctrl{1}		&  \targ{}		& 						& 			&			& 			& 			&			&			& \targ{}& \ctrl{1}	& \\
		\ket{\phi_3}			& \targ{}		&  				& 						& 			&			& 			& 			& 			&			& &\targ{}  	& \\
		\ket{\phi_4} 			& \ctrl{1}		&  \ctrl{2} 	& \targ{}				& 			&			& 			& 			&			&\targ{} 	& \ctrl{2} 	&  	\ctrl{1}	&\\
		\ket{\phi_5}			& \targ{}		&  				& 						& 			&			& 			&			& 			&			& &\targ{}		& \\	
		\ket{\phi_6} 			& \ctrl{1}		&  \targ{} 	 	& 						& 			&			&  			& 			&			&			& \targ{}&\ctrl{1} 	& \\
		\ket{\phi_7}			& \targ{}		&  				& 				& 	&			& 			& 			& & &&\targ{}	& \\
		\ket{\phi_8}			& \ctrl{1}		&  	\ctrl{1}			& \ctrl{1}			& \targ{}	&			& 			& 			&\targ{}	&\ctrl{1} &\ctrl{1}		& 		\ctrl{1}	& 	\\
		\vdots		& \vdots		&   	\vdots		&  \vdots 	&			&			& 			& 			&			& \vdots 	& 	\vdots	&		\vdots 		& \\
		\ket{\phi_{\lambda-2}}	& \ctrl{1}		&   			& 						& 			&			&			&			&  			& & & \ctrl{1}		& \\
		\ket{\phi_{\lambda-1}}	& \targ{}		&  				& 						& 			&			& 			& 			& 			& &	& \targ{} 		&
	\end{quantikz} 
	\caption{Quantum fanout in logarithmic depth.}
	\label{fig:fanout-tree}
\end{figure}
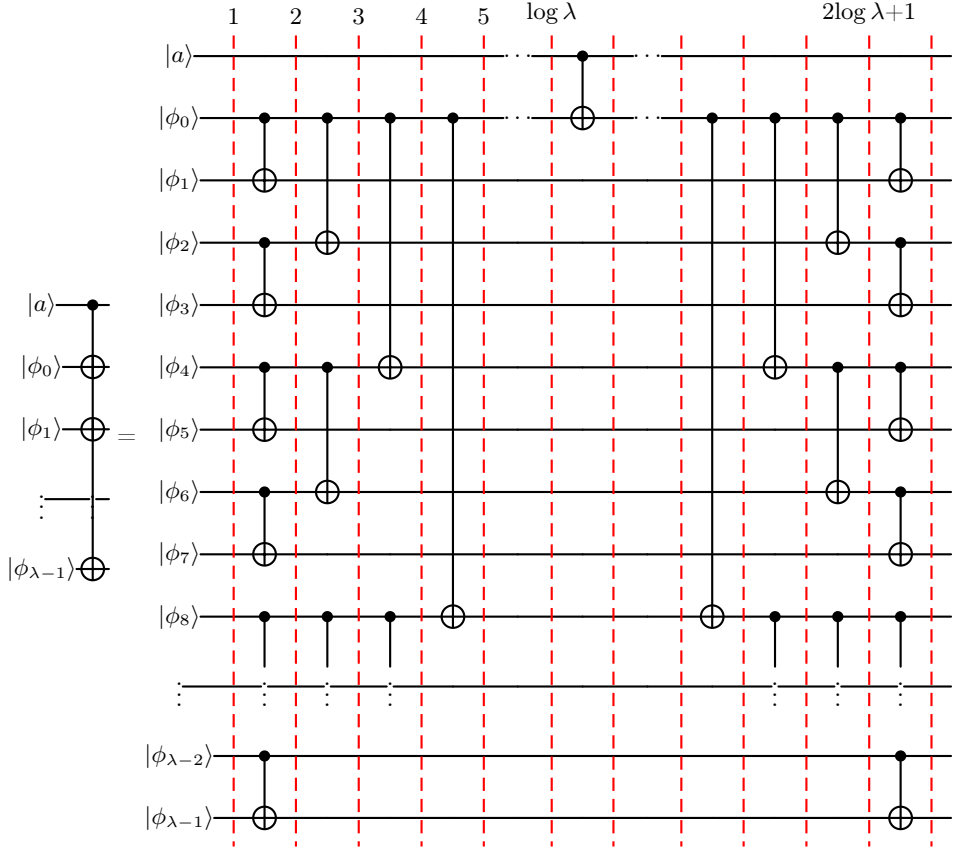

A direct circuit implementation of $\hat{U}_S$ presented in Fig.~\ref{fig:DA-select} consumes $n_Q(S) = \log N + \beta + 1$ qubits, $2N \cdot C_{\log N}X$ gates and $N\cdot CX^{\beta}$ gates. Here $C_nX^m$ denotes n-controlled m-NOT gate. The multi-qubit gates appearing in the $\hat{U}_S$ operator are decomposed into the Toffoli gates, which then can be executed on a fault-tolerant quantum device with a given number of Clifford+T gates. Three popular Toffoli-to-Clifford+T conversions are listed below:
\begin{enumerate}
	\item Exact: $1Toff = 7T + 6CX + 2H$
	\item Exact with ancilla: $1Toff = 4T+3CX+3Clifford + 1anc$
	\item Conditional: $1Toff =  4T+3CX+3Clifford$ when, a) phase of the target state is irrelevant; b) uncompute with the Toffoli gate follows.
\end{enumerate}
Improvements in the T-gate count are possible. Implementation of Gidney~\cite{Gidney2018} assumes $\log N-1$ additional ancilla qubits and a cost of $2\log N-2$ T-gates per each $C_{\log N}X$ gate giving the total of $n_T(S) = 4N\log N - 4N$ T-gates plus $N\beta+y\cdot (4\log N-4)$ Clifford gates, where $y$ denotes the number of Clifford gates per single Toffoli gate implementation. Ref.~\cite{Gidney2018} gives $y=6$. 
Cancellation techniques presented in Babbush et al.~\cite{Babbush2018} further reduce the T-gate cost of $\hat{U}_S$ to $n_T(S) =\mathcal{O}(N)$ at an extra cost of $\log N$ clean ancilla qubits. The $\cdot CX^{\beta}$ gate, or the \textit{quantum fanout}, can be implemented with depth-$\beta$ using $\beta$ CX gates. Quantum fanout can be implement in logarithmic depth by doubling the number of CX gates and arranging them in a tree-structure, as shown in Fig.~\ref{fig:fanout-tree}. Circuit depth for quantum fanout obeys the recurrence relation: $D_{fanout}(n) = D_{fanout}(n/2) + 2$, which leads to a $2\beta-1$ CX gate implementation in $D_{fanout}(\beta) = 2\log\beta+1$ depth, totalling to $(2N\beta - N)$ CX gates at $(2N\log\beta+N)$ depth. 
Note that the quantum fanout depth $D_{depth}$ is an upper bound, as some of the elements in $D(j)$ are zero. Whenever a $D(j,k)$ element is zero, the respective $CX$ gate is not present in the decomposition. 

In summary, quantum resources required to implement the $\hat{U}_S$-based implementation of the data-access oracle scale as follows 
\begin{eqnarray}
	n_T(S) =\mathcal{O}(N)\\
	n_C(S) = \mathcal{O}(N\beta)+\mathcal{O}(\log N) \\
	n_Q(S) = 2\log N + \beta
	\label{eq:DA-cost}
\end{eqnarray}

\begin{figure}\begin{tikzpicture}
		\node[scale=0.7] { 
			\begin{quantikz}[column sep=0.2cm]
				\ket{q}^{\otimes nQq}  &  \measure{j=?} & \\
				\ket{0} 				& \ctrl{1}\wire[u]{q}& \\
				\ket{0}^{\otimes \beta}	& \gate{U_s} &
			\end{quantikz}
			$\equiv$\begin{quantikz}[column sep=0.01cm]
				\ket{q}^{\otimes nQq}  &  \measure{q=0} 			& 					& \measure{q=0} 			& 	\measure{q=1} &			&\measure{q=1} 	& 						& \ldots & &\measure{q=\frac{N}{\lfloor \lambda \rfloor}} &&&&&& \ket{q}  \\
				\ket{r}^{\otimes nQr}  &  &&	& 	&	& 	& & & &  &\measure{r=0} 	& \measure{r=1} 	&\ldots & \measure{r=\lambda -1} &	& \ket{r} \\	
				\ket{0} 				& \targ{} \wire[u][2]{q} 	& \ctrl{1}			& \targ{} \wire[u][2]{q} 	& \targ{} \wire[u][2]{q}&\ctrl{1}		&  \targ{} \wire[u][2]{q} &&\ldots  	& \ctrl{1}&  \targ{} \wire[uu]{q}   &&&&&&\ket{0}\\
				\ket{0}^{\otimes \beta}	& 						& \gate{X^{B(0,0)}}\wire[u]{q} 	& 					&	& \gate{X^{B(1,0)}}\wire[u]{q}  	& && \ldots & \gate{X^{B(q_{max},0)}}\wire[u]{q}  &  &\gate[5]{\hat{U}_p(0)}\wire[u][2]{q} & \gate[5]{\hat{U}_p(1)}\wire[u][2]{q}& \ldots & \gate[5]{\hat{U}_p(\lambda-1)}\wire[u][2]{q} &&\ket{D(j)}\\
				\ket{0}^{\otimes \beta}	& 						& \gate{X^{B(0,1)}}\wire[u]{q} 	& 						&& \gate{X^{B(1,1)}}\wire[u]{q}  	& && \ldots & \gate{X^{B(q_{max},1)}} \wire[u]{q} & && &\ldots&&&\ket{\phi_{0}}\\
				\ket{0}^{\otimes \beta}	& 						& \gate{X^{B(0,2)}}\wire[u]{q} 	& 						&& \gate{X^{B(1,2)}}\wire[u]{q}  	& && \ldots & \gate{X^{B(q_{max},2)}}\wire[u]{q}  & &&& \ldots&&&\ket{\phi_{1}}\\
				\ket{0}^{\otimes \beta}	& 						& \vdots 	& 					&	& \vdots 	& && &\vdots & &&& \ldots &&& \vdots\\	
				\ket{0}^{\otimes \beta}	& 						& \gate{X^{B(0,\lambda-1)}}\wire[u]{q} 	& 						&& \gate{X^{B(1,\lambda-1)}} \wire[u]{q} 	& && \ldots & \gate{X^{B(q_{max},\lambda-1)}}\wire[u]{q}  & & & & \ldots &&& \ket{\phi_{\lambda-2}}
		\end{quantikz}};
	\end{tikzpicture}
	\caption{Schematic representation of the SELECTSWAP network of Low et al. $\ket{j}=\ket{q}\ket{r}\equiv \ket{\frac{N}{\lfloor \lambda \rfloor}}\ket{j\mod\lambda}$. $\ket{\phi}$ represent irrelevant states, carrying certain probability of measuring a subset of memory states. This probability and composition of the irrelevant states is determined by the $\hat{U}_S$ and $\hat{U}_P$ operators.}
	\label{fig:DA-selectswap}
\end{figure}
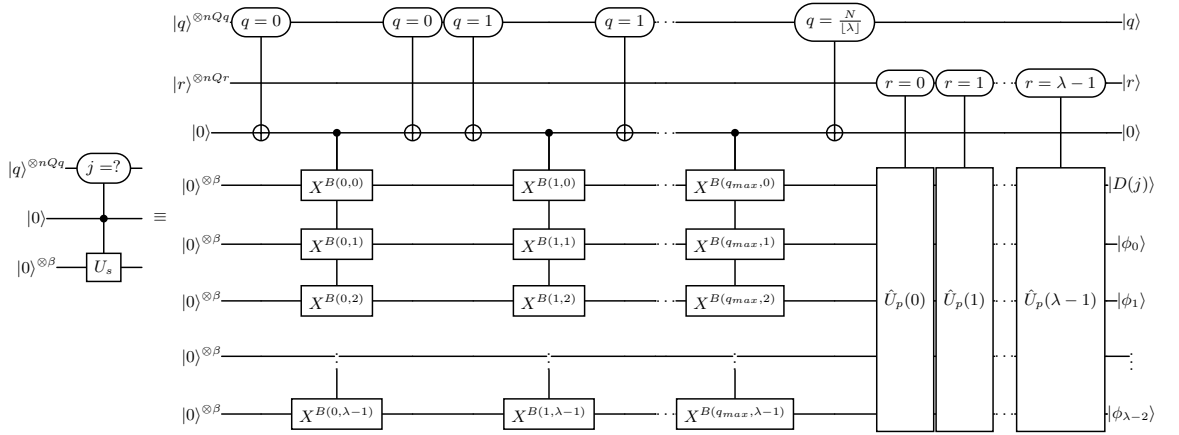

\paragraph{Ancilla-assisted data-access oracle}
The $\hat{U}_S$-based implementation of the data-access oracle can be improved by multiplexing target registers $\lambda$-times and applying appropriately tailored qubit swapping circuit, named the SELECTSWAP network in Low et al. The SELECTSWAP procedure realizes the transformation:
\begin{equation}
	\hat{D}\ket{j}\ket{0}\ket{\phi}=\ket{j}\ket{0}\ket{D(j)}\ket{G(j)}
	\label{DA:SELSWAP}
\end{equation}
where $\ket{G(j)}$ represents a $\beta(\lambda-1) $ qubit register and $\ket{D(j)}$ represents a $\beta$ qubit register. First, the input data register $\ket{j}$ is decomposed into the integer part of division by $\lambda$, $\ket{q}$, where $q=0,1,2,...,\lfloor\frac{N}{\lambda}\rfloor$, and the reminder $\ket{r}=\ket{j\mod \lambda}$, such that $j=q\lambda+r$. 
Controlled on the integer part $\ket{q}$, the quantum fanout is applied, as shown in Fig.~\ref{fig:DA-select}, to the multiplexed registers $\ket{0}^{\beta\lambda}$.  
The integer part register $\ket{q}$ requires $nQq = \lceil \log\lfloor\frac{N}{\lambda}\rfloor \rceil$ qubits whereas the reminder register $\ket{r}$ requires $nQr = \lceil \log \lambda \rceil$ qubits.
The value of the $\ket{q}$ register specifies the column and the reminder $\ket{r}$ specifies the row of the \textit{bitflip matrix} element returned by the oracle. It is assumed that $\ket{G(j)}$ is irrelevant to the calculation.
The uncertainty in the returned element of the bitflip matrix is removed by moving the marked reminder index to the topmost register, i.e. permute the rows of the bitflip matrix, written as:
\begin{equation}
	\hat{U}_P\ket{r}\otimes\ket{D(q,0)}\otimes\ket{D(q,1)}\otimes...\otimes\ket{D(q,\lambda-1)}=\ket{r}\otimes\ket{D(q,r)}\otimes\ket{D(q,\pi(1))}\otimes...\otimes\ket{D(q,\pi(\lambda-1))}
	\label{eq:SWAP}
\end{equation}
where $\pi(k)$ represents permutation defined by $\hat{U}_P$. A network of controlled-SWAP operations can realize the transformation shown in eq.~\ref{eq:SWAP}. 
This design leads to a decomposition of the data-access oracle given below as
\begin{eqnarray}
	\hat{D} = \hat{U}_P\hat{U}_S
	\label{DA:decomposition}
\end{eqnarray}
where $\hat{U}_P$ stands for the \textit{permutation} operators, i.e. the network of quabit swap operations depicted in Fig.~\ref{fig:DA-selectswap}. The selection and permutation operators can be written as:
\begin{eqnarray}
	\hat{U}_S=\sum_{q=0}^{\lfloor\frac{N}{\lambda}\rfloor}\ket{q}\bra{q}\otimes\prod_{k=0}^{\lambda-1}\hat{X}^{B_{\lambda}(q,k)}\\
	\hat{U}_P=\sum_{r=0}^{\lambda-1}\ket{r}\bra{r}\otimes\prod_{k=0}^{k_{max}}CSWAP\left(\pi(k),\sigma(k)\right)
	\label{eq:Up}
\end{eqnarray}
and the bitflip matrix generators $B_{\lambda}(q,k)$ can be written as:
\begin{eqnarray}
	B_{\lambda}(j,k) = b(q(j,\lambda)+b'(r(k))
	\label{eq:bitflipgen}
\end{eqnarray}
where $q(j,\lambda)=\lfloor\frac{j}{\lambda}\rfloor$ and $r(k)=j\mod \lambda$. 
Swapping respective entry registers $\ket{D(q,k)} \leftrightarrow \ket{D(q,k')}$ is realized bitwise by a sequence of $\beta$ controlled SWAP operations. 

The cost of the select part controlled on the $\ket{q}$ register is $\mathcal{C}(q) = 2 \frac{N}{\lfloor \lambda \rfloor} \mathcal{C}(C_{nQq}X)+\frac{N}{\lfloor \lambda \rfloor}\mathcal{C}(CX^{\lambda\beta})$. 
We employ the linear depth version of the quantum fanout resulting in the final gate count for the selection part:
$\mathcal{C}(q) =\mathcal{C}_T(q)+\mathcal{C}_{Cliff}(q) =  (4N\log N - 4N) + N\beta+6\cdot (4\log N-4)$ T-gates plus Clifford gates respectively. Note that $\mathcal{O}(N)$ T-gate implementation of Babbush et al., is possible~\cite{Babbush2018} at an extra clean qubit cost. 
The permutation part is controlled on the reminder register $\ket{r}$ with associated cost of $\mathcal{C}(r)=\lambda\beta \mathcal{C}(C_2X) +2\lambda\beta\mathcal{C}(CX)$.
Note that one controlled-SWAP (Fredkin) gate is equivalent to $2 CX$ gates and $1$ Toffoli gate. The Toffoli gate is considered equivalent to $7$ T-gates and $8$ Clifford gates, unless special conditions apply. The final cost of the permutation operator is thus   $\mathcal{C}(r)= \mathcal{C}_T(r)+ \mathcal{C}_{Cliff}(r) = 7\lambda\beta +10\lambda\beta $ T-gates and Clifford gates, respectively. Following Low et al.~\cite{Low2018}, is it possible to execute the permutation part in logarithmic depth with respect to memory register size $\lambda\beta$, at a cost of doubling the T-gate count. Here we assimilate the idea of minimizing the T-gate count and leave the $\mathcal{O}(\lambda\beta)$ Clifford depth.
In summary, the total cost for the SELECTSWAP network is given as
\begin{eqnarray}
	n_T(SP) = \mathcal{O}\left(\frac{N}{\lambda}\log\frac{N}{\lambda}+\lambda\cdot\beta\log\lambda\right)\\
	n_C(SP) =  \mathcal{O}\left(N\beta\right)+\mathcal{O}\left(\lambda\beta\right)\\
	n_Q(SP) = \lambda\beta + \lceil \log \frac{N}{\lfloor\lambda\rfloor} \rceil + \lceil \log \lambda \rceil + 1
	\label{eq:DA-SELSWAP-cost}
\end{eqnarray}
$\lambda = \sqrt{\frac{N}{\beta}}$ minimizes the T-gate count giving  $\tilde{n}_T(SP)= \tilde{\mathcal{O}}\left(\sqrt{N\beta}\right)$ where we assumed that $N>\beta$, which is satisfied for nearly all applications of the data-access oracle. $\tilde{\mathcal{O}}$ denotes asymptotic complexity with logarithmic terms dropped. The optimal Clifford gate count is then  $\tilde{n}_C(SP)=\mathcal{O}\left(N\beta\right)+\mathcal{O}\left(\sqrt{N\beta}\right)$ and the number of qubits $	\tilde{n}_Q(SP) = \sqrt{N\beta} + \frac{3}{2}\log N + \frac{1}{2}\log\beta  +1$.

One possible improvement of the above SELECTSWAP procedure is to exploit the structure of the Hamiltonian matrix. By sorting the Hamiltonian matrix elements by value we can arrange entries to ensure maximal length of the common bitmask between elements; neighbouring matrix elements in the permuted list have maximal possible common number of bits. This way we can apply identical operation to both entries appropriate number of times. 

The data-access oracle is used in state preparation based on the Hamiltonian coefficient and during controlled-rotation angle loading in the selection oracle for the Hamiltonian. State preparation can be achieved by numerous techniques. Here we choose the approach of Low et al.~\cite{Low2018}, originating in the works of Grover and Rudolph~\cite{grover2002}.

%\bibliography{beit.bib}

%

\onecolumngrid
\end{document}